\documentclass[aps,prx,twocolumn,showpacs,superscriptaddress,preprintnumbers,longbibliography]{revtex4-2} 
\usepackage{dcolumn}   
\usepackage{bm}        
\usepackage{amssymb}   
\usepackage{amsmath}
\usepackage{comment}
\usepackage{color}
\usepackage{slashed}
\usepackage{braket}
\usepackage{amssymb,amsmath,color}
\usepackage[caption=false]{subfig}
\usepackage{graphicx}  
\usepackage[export]{adjustbox}
\usepackage[colorlinks=true,citecolor=blue,linkcolor=red]{hyperref}
\usepackage{multirow}
\usepackage{rotating,booktabs}
\usepackage[verbose]{placeins}
\usepackage{mathrsfs}
\usepackage{stackengine}
\usepackage{soul}
\usepackage{float}

\newcommand{\Eq}[1]{Eq.~(\ref{#1})}

\newcommand{\Eqs}[2]{Eqs.~(\ref{#1}-\ref{#2})}
\newcommand{\Fig}[1]{Fig.~\ref{#1}}

\newcommand{\ua}{\mathord{\uparrow}}
\newcommand{\da}{\mathord{\downarrow}}

\newcommand{\TFI}{{TFI${}^2$}}
\usepackage[normalem]{ulem}
\newcommand{\delete}[1]{}

\makeatletter
\def\blfootnote{\xdef\@thefnmark{}\@footnotetext}
\makeatother
\begin{document}

\title{Stabilizer Scars}

\author{Jeremy Hartse}
\email{jhartse@uw.edu}
\affiliation{InQubator for Quantum Simulation (IQuS), Department of Physics, University of Washington, Seattle, WA 98195, USA.}
\author{Lukasz Fidkowski}
\affiliation{Department of Physics, University of Washington, Seattle, WA 98195, USA.}
\author{Niklas Mueller}
\affiliation{InQubator for Quantum Simulation (IQuS), Department of Physics, University of Washington, Seattle, WA 98195, USA.}
\affiliation{Center for Quantum Information and Control, University of New Mexico, Albuquerque, NM 87106, USA}
\affiliation{Department of Physics and Astronomy, University of New Mexico, Albuquerque, NM 87106, USA} 
\preprint{IQuS@UW-21-093}

\begin{abstract}
Quantum many-body scars are eigenstates in non-integrable isolated quantum systems that defy typical thermalization paradigms, violating the eigenstate thermalization hypothesis and quantum ergodicity. We identify exact analytic scar solutions in a $2+1$ dimensional lattice gauge theory in a quasi-1d limit as zero-magic resource stabilizer states. 
\end{abstract}
\maketitle

\noindent

\emph{Introduction}--- 
Unprecedented advances in controlling isolated quantum systems~\cite{gross2017quantum,browaeys2020many,altman2021quantum} have enabled the study of non-equilibrium phenomena in complex many-body systems, including thermalization, a pivotal subject in many fields~\cite{eisert2015quantum,d2016quantum,ueda2020quantum,berges2021qcd}.
According to the eigenstate thermalization hypothesis (ETH)~\cite{deutsch1991quantum,srednicki1994chaos}, non-integrable isolated systems thermalize, with quantum ergodicity ensuring that states explore a vast Hilbert space regardless of the initial state.

However, recent experiments have identified  so-called~\textit{quantum many-body scars} (QMBS),
where quantum ergodicity and the ETH are violated despite the system being strongly coupled and non-integrable~\cite{bernien2017probing,turner2018weak,serbyn2021quantum,moudgalya2022quantum,chandran2023quantum}. This has sparked significant theoretical and experimental interest~\cite{srivatsa2009quantum,turner2018quantum,moudgalya2018exact,moudgalya2018entanglement,choi2019emergent,shiraishi2019connection,schecter2019weak,iadecola2019exact,ok2019topological,chattopadhyay2020quantum,shibata2020onsager,lee2020exact,iadecola2020quantum,van2020quantum,pakrouski2020many,lin2020quantum,mark2020unified,moudgalya2020large,mondragon2021fate,wildeboer2021topological,windt2022squeezing,yao2022quantum,chen2022error,sun2023majorana,langlett2022rainbow,pizzi2024quantum,osborne2024quantum,budde2024quantum,calajo2024quantum,desaules2023prominent}, but  many questions are still open: For instance, the analytic mechanisms behind QMBS,  their stability in the thermodynamic limit and their fate as generic phenomena, beyond finely-engineered synthetic quantum systems, or use in quantum information and computing, are presently unclear.

Seeking insight into the mechanisms behind QMBS, in this Letter we investigate a quasi-$1$d limit of a strongly coupled and non-integrable $2+1$d model of a lattice gauge theory (LGT). LGTs are an important target for quantum simulators and computers because of their relevance in high-energy and nuclear physics~\cite{Banuls:2019bmf,Klco:2021lap,bauer2023quantum,beck2023quantum,bauer2023quantuma}, where thermalization is a central objective~\cite{mueller2022thermalization,zhou2022thermalization,ebner2024eigenstate, mueller2024quantum},  for topological phases~\cite{fradkin2013field,kleinert1989gauge}, or  quantum error correction~\cite{sarma2006topological,nayak2008non,lahtinen2017short}. 

\begin{figure}[t!]
    \centering
    \includegraphics[width=1.0\linewidth]{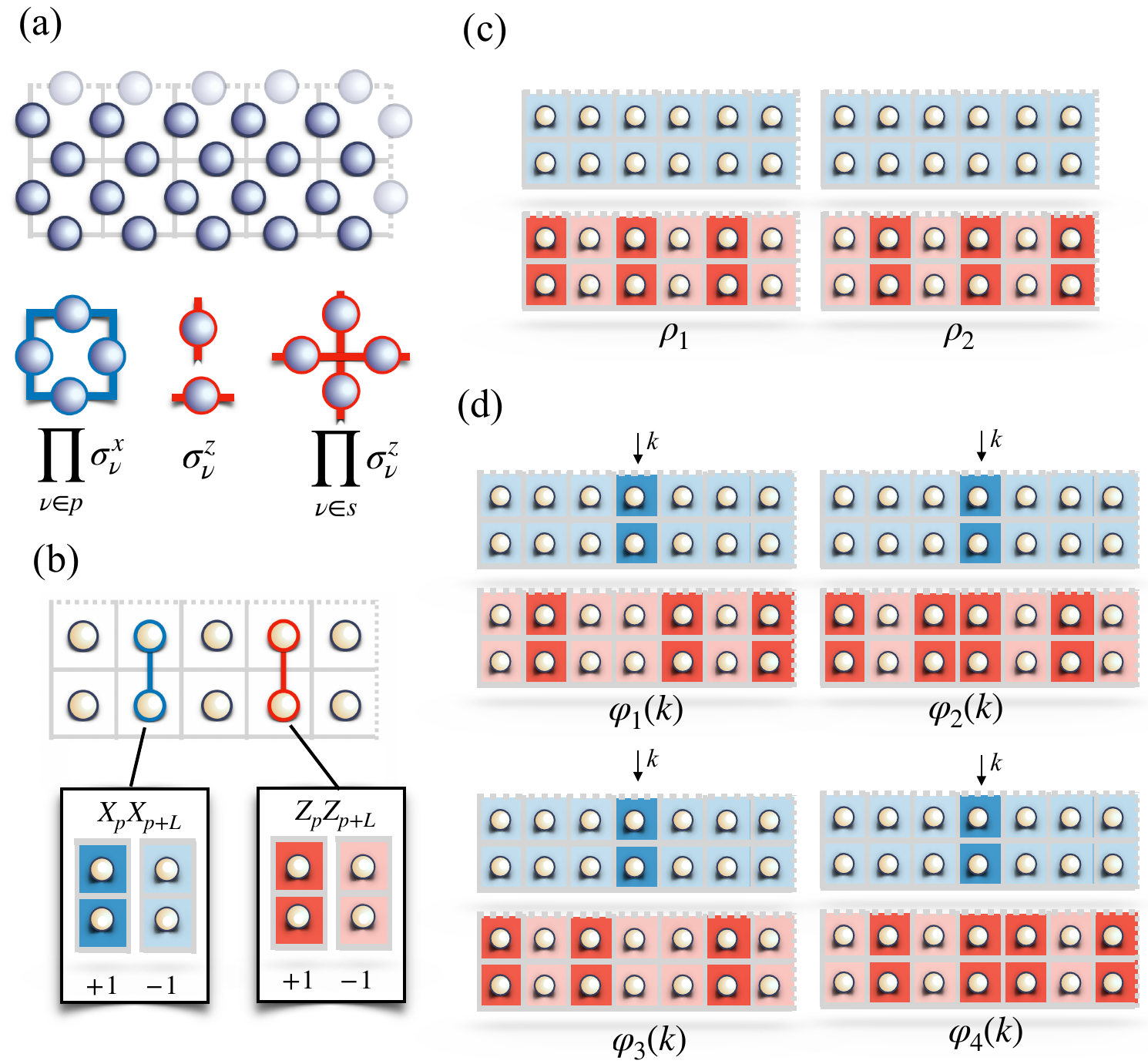}
    \caption{\textit{Quantum Many-Body Scars in $\mathbb{Z}_2$ LGT.} (a) Degrees of freedom of $\mathbb{Z}_2$ LGT on the edges of a (periodic) two-dimensional square lattice; plaquette, electric field and Gauss law operators are shown. (b) Illustration of Dual Ising model with stabilizers $X_pX_{p+L}$  (eigenvalues in light and dark blue) and $Z_pZ_{p+L}$ (eigenvalues in light and dark red). (c) QMBS scar solutions of a $L\times 2$ lattice where $L$ is even, shown in terms of $X_pX_{p+L}$ and $Z_pZ_{p+L}$ eigenvalues. (d) The scar subspace for odd $L$ is spanned by the states shown.} 
    \label{fig:overview}
\end{figure}

Our study, based on an Ising-LGT duality and fermionization,  uncovers exact analytic scar solutions for arbitrary couplings and in the thermodynamic limit. Importantly, the scar subspace of the model is spanned by certain \textit{stabilizer states}~\cite{gottesman1998heisenberg,aaronson2004improved} with zero magic resource ~\cite{bravyi2005universal,campbell2017roads},  revealing a connection between
complexity and quantum ergodicity.

\begin{figure*}[t]
    \centering
    \includegraphics[width=0.97\linewidth]{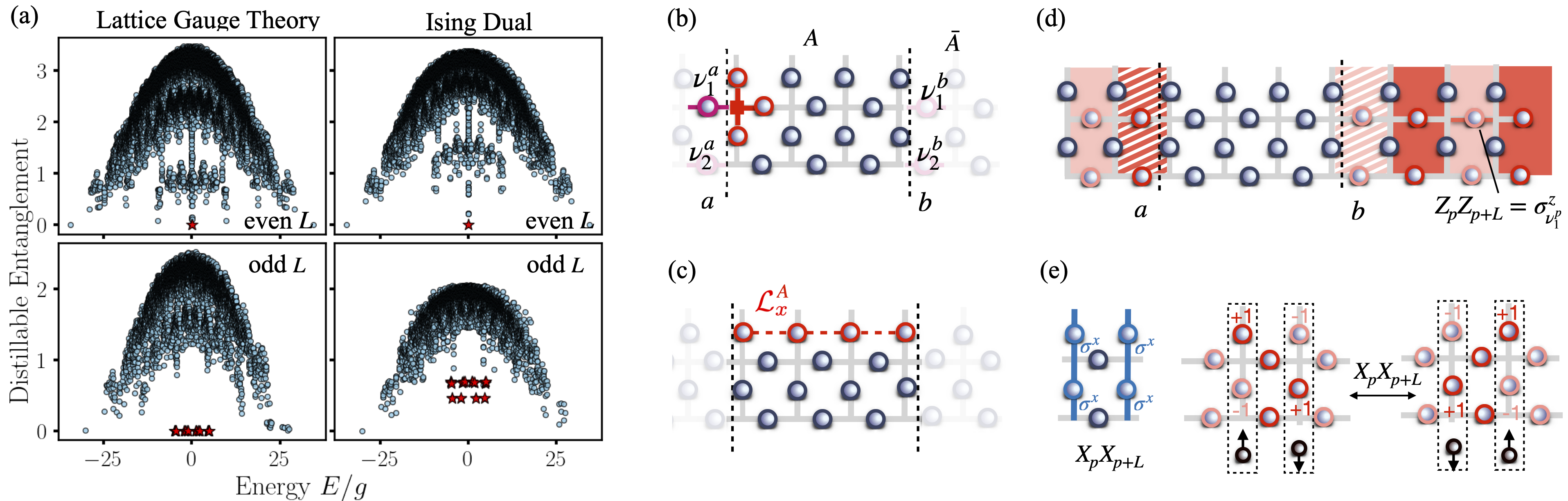}
    \caption{\textit{Quantum Many-Body Scars: Lattice Gauge Theory vs. Ising Dual.} (a) Top row: Distillable entanglement of all eigenstates for even $L$, for $L=8$ and $g=0.9$. The scar states, given by \Eq{eq:scarevenL1} and shown as red stars, appear in both the LGT (left panel) and the dual Ising model (right panel) and are characterized by exactly zero distillable entanglement. Bottom row: For odd $L$ ($L=7$ and $g=0.9$), the scar eigenstates form linear superpositions of the stabilizer states, \Eq{eq:LGTscars}. The  scars exhibit zero distillable entanglement in the LGT, but not in the dual Ising model. (b) Superselection sectors, $\sigma^z_\nu = \pm 1$ on each boundary link $\nu\in\{ \nu^a_1,\nu^a_2,\nu^b_1,\nu^b_2\}$, are associated with Gauss laws. (c)  An additional superselection sector, $\prod_{\nu \sim {\cal L}_x^A} \sigma^z_\nu \sim Z_a Z_b$, where $\scriptsize{\cal L}_x^A = {\cal L}_x \cap A$, arises from restricting $\prod_{\nu \sim {\cal L}_x} \sigma^z_\nu = V_x$ to $A$  (similar for $\bar{A}$). (d) The $Z_p Z_{p+L}$ eigenvalue configuration, for $p \leq a$ and $p \geq b$, is shown for the state $|\psi_{+1,-1}\rangle$ from \Eq{eq:plusminus}, with a defect trapped inside $A$. Along with the $X_pX_{p+L}$ and Gauss law constraints, the reduced density matrix of $|\psi_{+1,-1}\rangle$ on $\bar{A}$ (and $A$) has at most rank $2$. The light red and dark red colors indicate the $Z_p Z_{p+L}$ eigenvalue as in fig \ref{fig:overview}. (e) While $Z_p Z_{p+L} $ constrains horizontal links in the LGT, Gauss law eigenstates require vertically separated vertical links to be in either a $+1, -1$ or $-1, +1$ configuration, providing an effective spin-$1/2$ degree of freedom. Imposing the $X_p X_{p+L}$ constraint leads to two states—one even, one odd under $\prod_{\nu \sim {\cal{L}}_x^{\bar{A}}} \sigma^z_\nu$—resembling the degenerate ground state of a 1D Ising model without a transverse field. 
    } 
    \label{fig:scars}
\end{figure*}

\emph{Model}--- 
We consider $\mathbb{Z}_2$ LGT~\cite{wegner1971duality,horn1979hamiltonian}  with Hamiltonian,
\begin{align}\label{eq:Z2Hamiltonian}
    H=-\sum_{p}\prod_{\nu \in p}\sigma^x_{\nu}-g\sum_{\nu}\sigma_\nu^z
\end{align}
where $\sigma^{x/z}$ are Pauli matrices, $\nu$ are the links of a square lattice in $2+1$ dimensions with periodic boundary conditions, and $p$ are plaquettes, see ~\Fig{fig:overview}(a). Local Gauss law operators, $G_s\equiv \prod_{\nu \in s}\sigma^z_\nu$, where $\nu \in s$ are links $\nu$ originating from a lattice site $s$, see~\Fig{fig:overview}(a), and two ribbon operators $V_{x,y}\equiv \prod_{\nu \in \mathcal{L}_{x/y}}\sigma^z_{\nu}$, where $\mathcal{L}_{x/y}$ wind periodically around $x/y$, define superselection sectors, $[H,G_s]=[H,V_x]=[H,V_y]=0$~\cite{sachdev2018topological}.  We consider a lattice of $L\times 2$ plaquettes and, w.l.o.g., set $V_x=V_y=1$, and consider the combined $+1$ eigenspace of all $G_s$; see more in Supplemental Material \ref{app:entanglement}. 

The model described by \Eq{eq:Z2Hamiltonian} is dual to a $(2+1)$d transverse-field Ising model with periodic boundary conditions when restricted to the parity-even subspace, $\prod_{p}X_p=1$,  via the map $X_p\equiv\prod_{\nu \in p}\sigma^x_{\nu}$ and $Z_{p}Z_{p+\hat{a}}=\sigma_{\nu}^z$, where $p$ and ${p+\hat{a}}$ label the plaquettes adjacent to the link $\nu$, 
\begin{align}\label{eq:IsingDualHamiltonian}
    H^{\rm dual}= &- \sum_{p=1}^L [X_p+X_{p+L}] - \tilde{g}\sum_{p=1}^L Z_pZ_{p+L} \nonumber\\
     &-\sum_{p=1}^L g_p [Z_p Z_{p+1} + Z_{p+L} Z_{p+L+1} ]\,,
\end{align}
where $\tilde{g}\equiv g(1+V_y)$, and $g_p\equiv g$ for $p<L$ and $g_p=V_x g$ for $p=L$; $X_p$ and $Z_p$ are Pauli matrices centered on plaquettes, see~\Fig{fig:overview}(b). Our notation is such that plaquettes are numbered $1$ to $L$ from left to right in the top row and from $L+1$ to $2L$ from left to right in the bottom row. For more details about the duality see the Supplemental Material \ref{subsec:dual}.

\emph{Quantum Many-Body Scars}--- For an arbitrary value of the coupling $g$, and $L\times 2$ geometry, the model's eigenspace possesses a subspace of exact QMBS. For even $L$, this subspace is two-dimensional and spanned by the states,
\begin{align}\label{eq:scarevenL1}
    \rho_{ 1/2}\equiv  \prod_{p=1}^L \big[\frac{1\pm(-1)^pZ_pZ_{p+L}}{2} \big]\big[ \frac{1-X_pX_{p+L}}{2}\big]
\end{align}
where $\rho_{ 1/2} \equiv | \Psi_{ 1/2} \rangle\langle \Psi_{ 1/2} |$  are (eigen-)states with exactly zero energy, $ H | \Psi_{ 1/2} \rangle=0$. In Supplemental Material \ref{app:derivation}, we derive \Eq{eq:scarevenL1}.

Importantly, \Eq{eq:scarevenL1} are \textit{stabilizer states},  specified by the eigenvalues of  $X_p X_{p+L}$ and $Z_{p} Z_{p+L}$. The corresponding scar states in LGT formulation are obtained upon replacing $X_p X_{p+L}=\prod_{\nu \in p}\sigma^x_\nu\sigma^x_{\nu+L}$ and $Z_{p} Z_{p+L} =\sigma^z_\nu$; additionally, on this side of the duality,  Gauss's laws (and $V_x$, $V_y$) are stabilizers, i.e., 
\begin{align}\label{eq:LGTscars}
    \rho_{1/2}\rightarrow \rho_{1/2}\,  \prod_s\big[\frac{1+G_s}{2}\big] \frac{1+ V_x}{2} \frac{1+ V_y}{2}\,,
\end{align}
where the l.h.s. is the Ising and the r.h.s. the LGT expression. For brevity, we also use $X_p X_{p+L}$ and $Z_{p} Z_{p+L}$ when discussing the LGT, referring to $\prod_{\nu \in p}\sigma^x_\nu\sigma^x_{\nu+L}$ and $\sigma^z_\nu$, respectively.

For odd $L$, the scar subspace is spanned by $4L$ stabilizer states, $\varphi_\alpha\equiv | \varphi_\alpha \rangle \langle \varphi_\alpha|$,
\begin{align}\label{eq:scaroddL1}
    \varphi_\alpha(k) \equiv& 
  \prod_{q< k}\Big[\frac{1+s_\alpha (-1)^{k-q} Z_q Z_{q+L}}{2} \Big] \frac{1+t_\alpha Z_kZ_{k+L}}{2} \,  
    \nonumber\\
    \times &\prod_{q> k}\Big[\frac{1-s_\alpha(-1)^{k-q} Z_q Z_{q+L}}{2} \Big]\, 
   \Phi(k)\,.
\end{align}
Here, $\alpha=1,2,3,4$ and $k=1,\dots,L$; furthermore we abbreviate $s_1=s_4\equiv1$, $s_2=s_3\equiv-1$, as well as $t_2=t_4\equiv1$ and $t_1=t_3\equiv-1$. 
Additionally,
\begin{align}
    \Phi(k)\equiv   \prod_{q\neq k}\Big[\frac{1-X_q X_{q+L}}{2} \Big] \frac{1+X_kX_{k+L}}{2} \,.
\end{align}
Unlike for even $L$, \Eq{eq:scaroddL1} are not eigenstates, but rather they define a conserved subspace in which the action of $H$ maps scar states onto other scar states exactly while having zero overlap with non-scar states; we demonstrate this property in Supplemental Material \ref{app:derivation}. The states in the LGT formulation are defined analogously to~\Eq{eq:LGTscars}, with Gauss laws and $V_x$, $V_y$ additional stabilizers. In the \textit{End Matter} section we discuss the commutant algebra that generates the scar subspace.

\textit{Entanglement and Magic Resource--} The scar basis states, \Eq{eq:scarevenL1} and \Eq{eq:scaroddL1}, have zero magic resource because they are stabilizer states. We investigate their entanglement, focusing on the von Neumann entanglement entropy of a  bipartition along the long ($L$) lattice direction,
$ \rho_A = \text{Tr}_{\bar{A}}(\rho) $,
where $A$ is the subsystem and $\bar{A}$ its complement. The reduced density matrix $\rho_A\equiv \bigoplus_s \rho_{A}^{(s)}$ exhibits a block-diagonal structure, whose origin we explain below. As a result, the von Neumann entanglement entropy can be decomposed into distillable and symmetry  contributions,
\begin{align}\label{eq:vonNeumanncomponents}
    S_{\rm vN} = S_{\rm vN}^{\rm dist}-\sum_s p_s \log(p_s) 
    \,,
\end{align}
where the distillable von Neumann entropy is
\begin{align}\label{eq:disent}
    S_{\rm vN}^{\rm dist} = \sum_s p_s S_{\rm vN}^{(s)} \,, \quad S_{\rm vN}^{(s)} =-\text{Tr}
   \big( \tilde{\rho}_{A}^{(s)} \log(\tilde{\rho}_{A}^{(s)} ) \big)\,,
\end{align}
with $\tilde{\rho}_{A}^{(s)} \equiv {\rho}_{A}^{(s)} / p_s  $ and $p_s=\text{Tr}({\rho}_{A}^{(s)})$.
Although the LGT and Ising models are dual to each other, the symmetries of $\rho_A$ and its entanglement structure generally differ between the two formulations. 

In the top row of \Fig{fig:scars}(a) we show the distillable entanglement for even $L=8$ and $g=0.9$ in both the LGT (left panel) and the Ising model (right panel). For both zero-energy scar eigenstates, the distillable entanglement vanishes. In contrast, the bottom row illustrates the same quantity for \textit{odd} $L=7$. Here, the eigenstates are linear superpositions of the stabilizer scar solutions, \Eq{eq:scaroddL1}. In the LGT, the distillable entanglement remains zero for any coupling, whereas it is non-zero in the Ising dual.

 \textit{Zero distillable entanglement in lattice gauge theory scar states---} The conserved scar subspace exists in both the Ising model and the LGT, but the scars' distillable entanglement is \textit{exactly zero for any value of coupling} only in the LGT owing to the special superselection sector structure in the latter. To explain this structure, we consider the entanglement cut shown in \Fig{fig:scars}(b). Gauss' law implies that $\sigma^z_\nu = \pm 1$ on each boundary link $\nu\in\{ \nu^a_1,\nu^a_2,\nu^b_1,\nu^b_2\}$ is a  distinct superselection operator. However, the constraint $\prod_{\nu \sim {\cal L}_y} \sigma^z_\nu = V_y = 1$, where ${\cal L}_y$ is any non-trivial loop in the $y$ direction and $\nu \sim {\cal L}_y$ are the links cut by ${\cal L}_y$, gives ${\sigma^{ \scriptstyle z}}_{\scriptscriptstyle \nu_{\scriptscriptstyle 1}^{ \scriptscriptstyle a/b}} {\sigma^{ \scriptstyle z}}_{\scriptscriptstyle \nu_{\scriptscriptstyle 2}^{ \scriptscriptstyle a/b}} = 1$. Consequently, there is only one superselection operator ${\sigma^{ \scriptstyle z}}_{\scriptscriptstyle \nu_{\scriptscriptstyle 1}^{ \scriptscriptstyle a/b}}$ associated with each boundary, which in the Ising dual equal $Z_a Z_{a+L}$ and $Z_b Z_{b+L}$. A third superselection operator arises from the constraint $\prod_{\nu \sim {\cal L}_x} \sigma^z_\nu = V_x=1$. Its restriction to $A$, shown in \Fig{fig:scars}(c), $\prod_{\nu \sim {\cal L}_x^A} \sigma^z_\nu \sim Z_a Z_b$, where ${\cal L}_x^A = {\cal L}_x \cap A$, gives the third superselection operator. We refer to Supplemental Material \ref{app:entanglement} for more details.

These three superselection operators result in zero distillable entanglement for the scar energy eigenstates in the LGT (but not for the Ising dual): Using the notation of \Fig{fig:overview}(d), a scar energy eigenstate is a superposition 
\begin{align} \label{eq:superp}
    |\psi\rangle = \sum_{1\leq k\leq L, 1\leq i \leq 4} c_{k,i} |\phi_i(k)\rangle
\end{align}
where $\sum_{k,i} |c_{k,i}|^2=1$ and $|\phi_i(k)\rangle$ is a LGT scar basis state via our dictionary. We now group the terms in this superposition according to their boundary superselection sectors 
\begin{align}
    |\psi\rangle = \sum_{\alpha,\beta=\pm 1} |\psi_{\alpha,\beta}\rangle\,,
\end{align}
i.e., $Z_a Z_{a+L}|\psi_{\alpha,\beta}\rangle=\alpha|\psi_{\alpha,\beta}\rangle$ and $Z_b Z_{b+L}|\psi_{\alpha,\beta}\rangle=\beta|\psi_{\alpha,\beta}\rangle$. The key observation is that, when considering a reduced density matrix on $A$, the four terms fall into separate superselection sectors. 

 Since the argument is nearly identical for all four states $| \psi_{\alpha,\beta}\rangle$, we will demonstrate it for $\alpha=1$ and $\beta=-1$ and $a$ and $b$ are even. We consider
\begin{align} \label{eq:plusminus}
    |\psi_{+1,-1}\rangle &= \sum_{(a<k<b) \,\text{and}\, (k \,\text{even}), i=1,4} c_{k,i} |\phi_i(k)\rangle \nonumber \\ 
    &+\sum_{(a<k<b) \,\text{and}\, (k \,\text{odd}), i=2,3} c_{k,i} |\phi_i(k)\rangle \nonumber \\
    &+ c_{a,4}|\phi_4(a)\rangle + c_{b,1}|\phi_1(b)\rangle
\end{align}
which represents a state where the $Z_k Z_{k+L}$ defect, as illustrated in Fig. \ref{fig:overview}(d), is trapped within region $A$ -- the set-up is shown in \Fig{fig:scars}(d).  
Any state described by \Eq{eq:plusminus} shares the same stabilizer values for $X_p X_{p+L}$ and $Z_p Z_{p+L}$ for all $p \leq a$ and $p \geq b$~\footnote{By fixing the superselection sector, $\alpha$ and $\beta$, we create a state with a defect localized either inside or outside of $A$. A superposition with defect trapped \textit{outside} of $A$ would have all $X_p X_{p+L}$ and $Z_p Z_{p+L}$ identical everywhere \textit{inside} of $A$.}, which imposes $2(L-b+a-1)$ constraints. Additionally, enforcing Gauss’s law in the complement region introduces $2(L-b+1)-1$ further constraints. For given $\alpha$ and $\beta$, there are $4(L+a-b)-2$ qubits in the complement, leaving exactly one remaining spin degree of freedom. Consequently, the reduced density matrix of the state $|\psi_{+1,-1}\rangle$ of both the complement and $A$ has at most rank 2. Furthermore, as  explained in \Fig{fig:scars}(e), these two states differ in their eigenvalues of $\prod_{\nu \sim {\cal L}_x, \nu \in {\bar{A}}} \sigma^z_\nu$. As a result, the distillable entanglement of an arbitrary superposition is zero. 
\begin{figure}[t]
    \centering
    \includegraphics[width=0.92\linewidth]{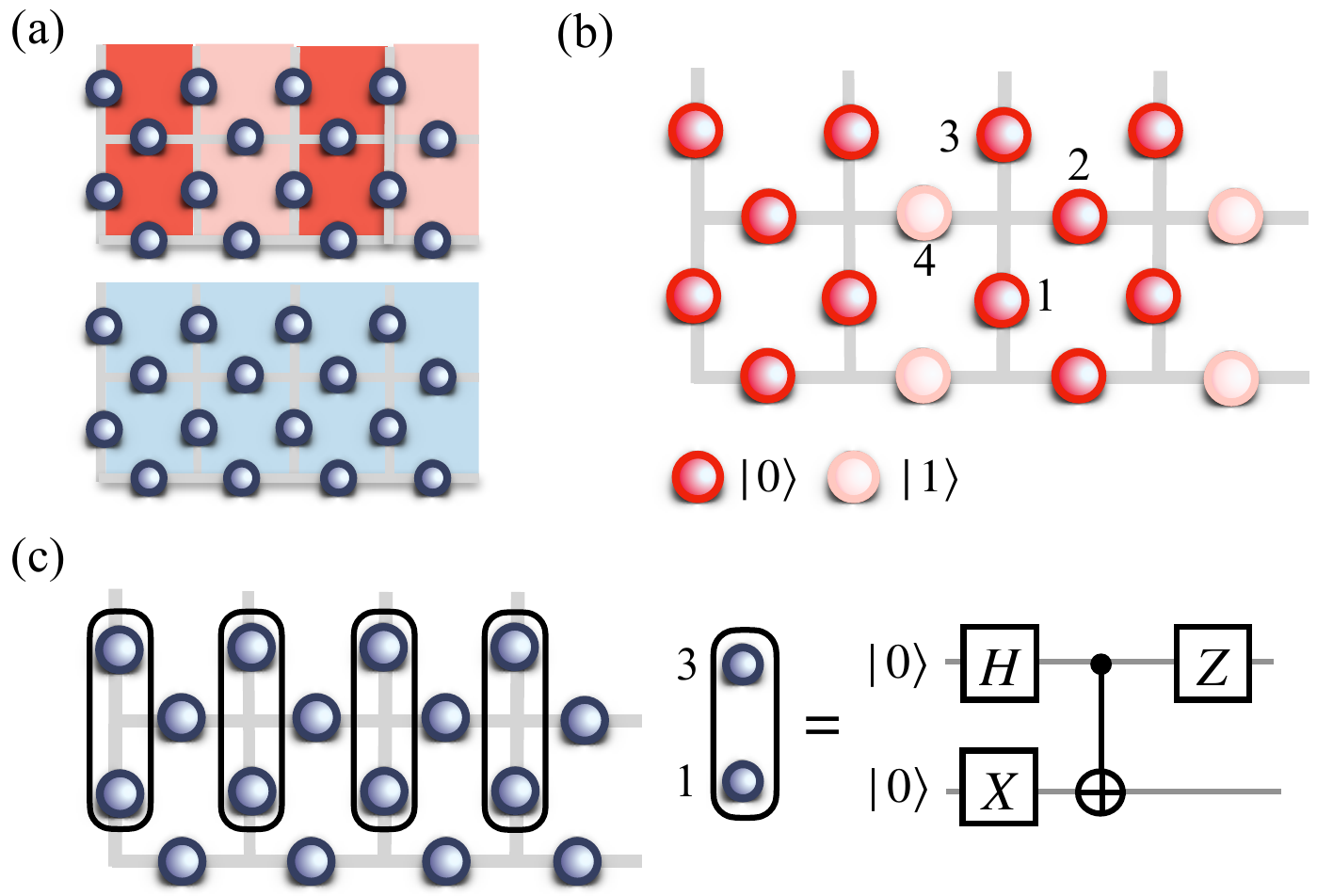}
    \caption{\textit{QMBS Preparation.} (a) Example of a scar basis state specified by the eigenvalues of $ X_p X_{p+L} $ and $ Z_p Z_{p+L} $. (b) Initial state for the preparation algorithm in the LGT formulation ($ V_y = 1 $). Horizontal links are initialized in $ \sigma^z $ (electric) eigenstates, while vertical links are initially set to $ |0\rangle $. (c) The state preparation circuit operates on pairs of vertically separated qubits located at $ \nu $ and $ \nu + \hat{y} $. An example is shown where the circuit prepares a state with $ \sigma^x_\nu \sigma^x_{\nu+\hat{y}} = -1 $ and $ \sigma^z_\nu \sigma^z_{\nu+\hat{y}} = -1 $; combined with the horizontal links, this yields a Gauss law +1 eigenstate. Minor circuit adjustments enable preparation of any combination of $ \sigma^x_\nu \sigma^x_{\nu+\hat{y}} = \pm1 $ and $ \sigma^z_\nu \sigma^z_{\nu+\hat{y}} = \pm1 $ states, allowing access to any scar subspace basis state. The circuit maintains product-state structure along $ L $; each state remains a 2-qubit product state.
    \label{fig:stateprep}}
\end{figure}

The only symmetry in the transverse Ising model is the global spin-flip symmetry, which gives rise to \textit{two} superselection sectors within which entangled superpositions of the scar states can exists, see \Fig{fig:scars}(a) for odd $L$.

\emph{Experimental Realization.}--- Finally, we propose an experimental implementation by time-evolving a scar stabilizer basis state in the LGT, versus preparing a non-scar state.  Importantly, since the scar subspace is exactly decoupled from the rest of the Hilbert space, an initial scar state is constrained to explore only a polynomially small subspace of $4L$ states, as opposed to a non-scar state, which may probe the exponentially large $2^{2L-1}$ dimensional Hilbert space.
Focusing on odd $L$, the state preparation protocol for a a simultaneous eigenstate of $Z_{p}Z_{p+L}$, $X_{p}X_{p+L}$, and the Gauss law operators is outlined in \Fig{fig:stateprep}, with \Fig{fig:stateprep}(a) showing a sample target state. This is achieved by configuring the horizontal and vertical links as depicted in \Fig{fig:stateprep}(b), followed by a the algorithm shown in \Fig{fig:stateprep}(c) which shows that the scar basis states are 2-qubit product states. Non-scar initial states are randomly drawn  ($z$-basis) product states that are consistent with Gauss law and have zero overlap with the scar subspace. 

We focus on two observables that distinctly differentiate scar dynamics from non-scar dynamics: the overlap between the initial and time-evolved state, known as the Loschmidt echo
\begin{align}
    L(t; |\psi_0\rangle)\equiv \langle \psi_0| \exp\{-iHt\} |\psi_0\rangle\,,
\end{align}
and the distillable von Neumann entropy defined in~\Eq{eq:disent}. While exact time evolution is performed here, in an experimental setting one would typically employ Trotterized time evolution---for an alternative scheme that avoids it see \ref{app:random_circuit}. 
\begin{figure}[t]
    \centering
    \includegraphics[width=0.68\linewidth]{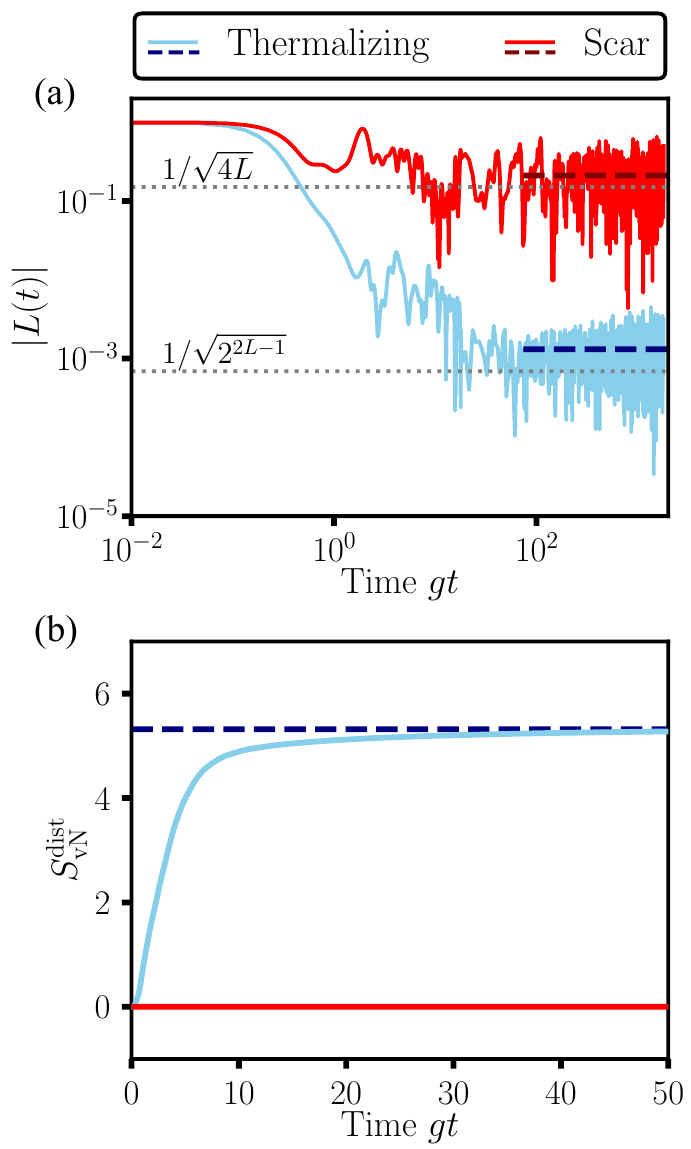}
    \caption{\textit{Experimental Realization.} (a) The absolute value of the Loschmidt echo, $ |L(t)| $, is shown as a function of time, comparing an initial state within the scar subspace (red lines) to a thermalizing state (blue lines) for system size $ L = 11$. The time-averaged value at late times is indicated by the dark red dashed line for the scar state. The scar echo decays as $ |L(t)| \sim 1/\sqrt{4L} $ (gray dotted line), where $4L$ is the size of the scar subspace. In contrast, the thermalizing state (blue lines) explores a much larger Hilbert space. For a random circuit that fully explores this space, $ |L(t)|$ would decay to $ |L(t)| \sim 1/\sqrt{2^{2L - 1}} $ (gray dotted lines), however, due to energy conservation, the thermalizing state's late-time average (dark blue dashed line) remains slightly above this value. (b)~The distillable entanglement $ S^{\rm dist}_{\rm vN} $ of a time-evolved scar state (red) remains zero throughout. By contrast, for a thermalizing state (light blue), $ S^{\rm dist}_{\rm vN} $ increases over time until reaching a constant, non-zero saturation value (blue dashed line).
    \label{fig:echo}}
\end{figure}

In \Fig{fig:echo}(a), we plot the absolute value $ |L(t)| $ as a function of time for one of the $ 4L $ scar basis states (solid red lines) for $L=11$.  For comparison, we also show the dynamics of a non-scar initial state (solid blue lines). The evolution of scar versus non-scar states differ dramatically. In the polynomially small scar subspace the Loschmidt echo decays to $|L(t)|\sim 1/\sqrt{4L}$. However, no full revivals are observed, as the energies of eigenstates in the scar subspace are not equidistant.
In contrast, non-scar initial states exhibit quantum ergodic behavior, where the initial state explores the (exponentially large) full Hilbert space that is available to it based on energy conservation~\footnote{This space is, roughly, the space spanned by eigenstates in the window $E\pm\Delta E/2$, where $E=\langle H \rangle$ and $\Delta E^2\equiv \langle H^2 \rangle - \langle H\rangle^2 $, and $\langle \cdot \rangle$ the initial state expectation value.}. Eventually, this state thermalizes, with the Loschmidt echo decaying to a much lower value. For comparison, we show  $|L(t)| \sim 1/\sqrt{2^{2L-1}}$, the saturation value in the absence of energy conservation.
Loschmidt echos can be measured in experiment using ancilla-based interferometry~\cite{somma2002simulating,knap2013probing,mueller2023quantum}.

\Fig{fig:echo}(b) shows the distillable entanglement, revealing an even more striking difference between scar and non-scar sectors. In the scar sector, the distillable entanglement remains exactly zero at all times, consistent with the argument presented earlier. In contrast, for states outside the scar subspace, the distillable entanglement increases over time, indicating that the system is thermalizing. 
Using random-measurement tools~\cite{elben2023randomized}, for the model at hand, (distillable) entanglement can be measured directly~\cite{bringewatt2024randomized}, or see \cite{mueller2024quantum} for an experimental demonstration via entanglement Hamiltonian tomography~\cite{kokail2021entanglement,kokail2021quantum}.

\emph{Conclusions}--- In this Letter, we identified exact quantum many-body scars in a quasi-1d limit of a 2+1 dimensional $\mathbb{Z}_2$ LGT, for arbitrary coupling, that are based on zero-magic resource stabilizer states.  An Ising-LGT duality and mapping onto a fermionic representation allowed us to analytically show their existence via a non-trivial commutant of the algebra generated by the Hamiltonian terms which we explicitly compute. Notably, in the gauge theory formulation,  scar states—and arbitrary superpositions—always have \textit{zero distillable entanglement}. The analytical construction presented in this manuscript does not extend beyond the $L \times 2$ geometry. The specific scar states we identified are absent in larger geometries such as $L \times 3$ and $L \times 4$. Although we have explored these cases numerically for a limited range of coupling values $g$, constrained by the exponentially increasing computational cost, we cannot rule out the existence of other, qualitatively different types of QMBS in these geometries.

Our finding reveals a  connection between computational complexity~\cite{liu2022many,tarabunga2023many,rattacaso2023stabilizer,catalano2024magic,robin2024magic,brokemeier2024quantum, chernyshev2024quantum} and quantum ergodicity, finding that states with zero distillable entanglement also possess low non-stabilizerness.Additionally, it indicates that identifying QMBS through entanglement may be unreliable: If we had computed the (total) von Neumann entropy, instead of the \textit{distillable} entanglement, we likely would not have initially detected these states. In contrast, while not explicitly computed,  the QMBS subspace can be distinguished from generic states, i.e., those that obey the ETH and exhibit quantum ergodicity, through measures of magic resource such as stabilizer Rényi entropies (SRE)~\cite{leone2022stabilizer}. In particular, we highlight~\cite{smith2024non}, which investigates SREs of scarred states in the PXP model.  Specifically, typical states have magic resource that scales \textit{linearly} with $L$~\cite{leone2022stabilizer,turkeshi2023pauli}, whereas the magic resource of states in the scar subspace is constrained by the superpositions that can be formed among the $4L$ zero-magic resource states. Based on arguments presented in~\cite{leone2022stabilizer}, we conjecture that the magic resource of states within the scar subspace is at most \textit{logarithmic in} $L$~\footnote{Our argument is based solely on the observation in~\cite{leone2022stabilizer} that, for a spin-$1/2$ (qubits) system, the $\alpha$-SRE is bounded by $\log(d)$ (specifically, it is bounded by $\log(d+1) - \log(2)$ for $\alpha = 2$), where $d$(=$4L$) is the dimension of the space. Although the scar subspace is not merely a system of qubits and the assumptions of~\cite{leone2022stabilizer} do not directly apply, we would find a different scaling with $L$ surprising}.

In the experimental realization we propose, classical simulability or intractability depends solely on the choice of the initial state, while the quantum circuit for time evolution remains fixed. This
distinction may be useful for verification and benchmarking, and will be pursued in a forthcoming manuscript.

\emph{Acknowledgements}---
We thank Marc Illa, Martin Savage, and Nikita Zemlevskiy for discussions. J.H. and N.M. acknowledge funding by the DOE, Office of Science, Office of Nuclear Physics, IQuS (\url{https://iqus.uw.edu}), via the program on Quantum Horizons: QIS Research and Innovation for Nuclear Science under Award DE-SC0020970. 
 L.F. is supported by NSF DMR-2300172.

\emph{Data availability}--- The data that support the findings of this Letter are openly available~\cite{hartse_2025}.

\bibliography{bibi.bib}

\section*{End Matter}
\emph{End Matter: Generating Algebra}--- Here we discuss the commutant algebra that generates the scar subspace. As is shown in Supplemental Material \ref{app:derivation}, the scar subspace is preserved not just by $H$, but also by the individual terms $X_p+X_{p+L}$, $Z_p Z_{p+1} + Z_{p+L} Z_{p+L+1}$, and $Z_p Z_{p+L}$ appearing in $H$.  Following \cite{Motrunich1,Motrunich2} we denote the algebra generated by these terms as $\cal{A}$, and its commutant - that is, the algebra of all operators that commute with all operators in $\cal{A}$ in the endomorphism algebra of the full Hilbert space - as $\cal{C}$. The existence of QMBS is reflected in a non-trivial $\cal{C}$ --  one can find a basis in which all elements of $\cal{C}$ share the same block diagonal form, and the scar subspace corresponds to a particular block. 

We compute the commutant $\cal{C}$ in the two leg transverse field Ising ladder by using a fermionic representation of the operator algebra introduced in the Supplemental Material \ref{app:derivation}. Here, we  give a brief overview of the argumentation: this  representation comes about from applying the ordinary Jordan-Wigner (JW) transformation from left to right along the first row (sites $1,\ldots, L$), and then along the second row (sites $L+1,\ldots, 2L$).  The two-level fermionic system at each site $p=1,\ldots, 2L$ can be described using two Majorana operators $\gamma_p, {\bar{\gamma}}_p$, which can then be recombined into complex fermions $c_p, c^\dagger_p$ and ${\bar c}_p, {\bar c}_p^\dagger$, $p=1,\ldots,L$ by $c_p = \frac{1}{2}\left(\gamma_p - i\gamma_{p+L}\right)$, ${\bar{c}}_p = \frac{1}{2}\left({\bar{\gamma}}_p - i{\bar{\gamma}}_{p+L}\right)$.  
The virtue of this representation is that two of the three types of terms generating $\cal{A}$ fermionize into manifestly particle number, ${\hat{N}}=\sum_{p=1}^L (c_p^\dagger c_p + {\bar{c}}_p^\dagger {\bar{c}}_p)$,  conserving operators quadratic in the fermions: $X_p+X_{p+L} \rightarrow 2i(c_p {\bar c}_p^\dagger + c_p^\dagger {\bar{c}}_p)$ and $Z_p Z_{p+1} + Z_{p+L} Z_{p+L+1} \rightarrow 2i( {\bar{c}}_p c_{p+1}^\dagger + {\bar{c}}_p^\dagger c_{p+1} )$ (for $p<L$).
As shown in Supplemental Material \ref{app:derivation}, the remaining terms, $ Z_p Z_{p+L} $ and $ Z_1 Z_L + Z_{L+1} Z_{2L} $, fermionize to operators that include a residual JW string, exhibiting a particle-hole symmetry that conjugates $ \hat{N} $ to $ 2L - \hat{N} $. This implies that $ (\hat{N} - L)^2 $ commutes with these terms and, therefore, with all of $ \mathcal{A}$, making it an element of $ \mathcal{C} $. In fact, $ (\hat{N} - L)^2 $ generates the entirety of $ \mathcal{C} $, as follows from Table II of~\cite{Motrunich3}.
In terms of the commutant $\cal{C}$, for the even $ L $ scar subspace the fermionic dual has overall even fermion parity, and the two scar states correspond to $({\hat{N}}-L)^2 = L^2$, i.e. the all empty $\hat{N}=0$ and the completely filled $\hat{N}=2L$ states. For odd $L$ the fermionic dual has odd fermion parity, and hence the least degenerate eigenspace of $({\hat{N}}-L)^2$ corresponds to the eigenvalue $(L-1)^2$, has dimension $4L$, and is spanned by the one-particle and one-hole states~\footnote{The scars identified here differ from those in Ref. \cite{Schoutens}, where an exact low-entanglement scar state, $|\psi'_{E=0}\rangle$, was constructed as a tensor product over `diagonal' singlets (Eq. 8 in Ref. \cite{Schoutens}). This state, while an eigenstate of the two-leg transverse field Ising model, lies outside the scar subspaces computed in this work.  In particular, the state $|\psi'_{E=0}\rangle$ is not robust against translation symmetry breaking in the $x$-direction, as opposed to the scar subspaces found in this work.}.

\section*{Supplemental Material}
\label{sec:sup}

\subsection{Lattice Gauge Theory and Ising Dual}
\label{subsec:dual}

In this subsection, we provide details and nomenclature of the $\mathbb{Z}_2$ LGT model that we work with, along with the mapping to a $(2+1)$d Ising model. The model described by \Eq{eq:Z2Hamiltonian} is well studied, as is the duality, and details can be found for instance in Refs.~\cite{wegner1971duality,horn1979hamiltonian,sachdev2018topological}.

\Eq{eq:Z2Hamiltonian} consists of a 'magnetic term' (or 'plaquette term'), $\sum_p\prod_{\nu \in p} \sigma^x_\nu$, and an 'electric term,' $\sum_\nu \sigma^z_\nu$. This nomenclature is borrowed from lattice quantum electrodynamics (LQED), to which the model is related when spin-$\frac{1}{2}$ degrees of freedom are generalized to $N$-dimensional ones, yielding a $\mathbb{Z}_N$ LGT. In the limit $N \to \infty$, this recovers LQED without matter. The model describes a \textit{gauge theory} because of the presence of local constraints, \textit{Gauss laws},
\begin{align}\label{eq:Glaw}
    G_s\equiv \prod_{\nu \in s}\sigma^z_\nu\,.
\end{align}
These operators commute with the Hamiltonian, i.e., $[H, G_s] = 0$ for all $s$, thereby defining superselection sectors. The sector where all $G_s = 1$ is typically referred to as the 'physical' sector. Continuing the analogy with electrodynamics, this is the sector without background charges, where electric field lines originate from (positive) electric charges—specifically, $\mathbb{Z}_2$ charges in our case~\cite{sachdev2018topological}.
For periodic boundary conditions, the model described by \Eq{eq:Z2Hamiltonian} has two additional superselection operators, $[H, V_x] = [H, V_y] = 0$, where $V_x$ and $V_y$ are defined in the main text as 'ribbons' of electric field operators that wind periodically around the $x$- and $y$-dimensions of the lattice, respectively~\cite{sachdev2018topological}.

The LGT-Ising duality was first described in~\cite{wegner1971duality}, and we refrain from discussing its details beyond those pertinent to our study. The precise mapping between the LGT and the Ising dual can be found in Table~\ref{tab:LGTdual}. In the Ising dual, the Gauss law constraints are eliminated, and $V_x$ and $V_y$ become parameters in the Hamiltonian (see \Eq{eq:IsingDualHamiltonian}), which take values $\pm 1$ depending on the superselection sector in the LGT. For periodic boundary conditions in the LGT, the product of all plaquette operators is unity by construction. On the dual side, this condition must be enforced by requiring $\prod_p X_p = 1$.

\begin{table}[t]
    \centering
    \begin{tabular}{c|cc|cc}
     \toprule \toprule
         & \multicolumn{2}{c|}{$\mathbb{Z}_2$ {LGT}} & \multicolumn{2}{c}{{Ising Dual}} \\
    \addlinespace
        \midrule
         {Hilbert space}  & \multicolumn{2}{c|}{ \includegraphics[width=2cm]{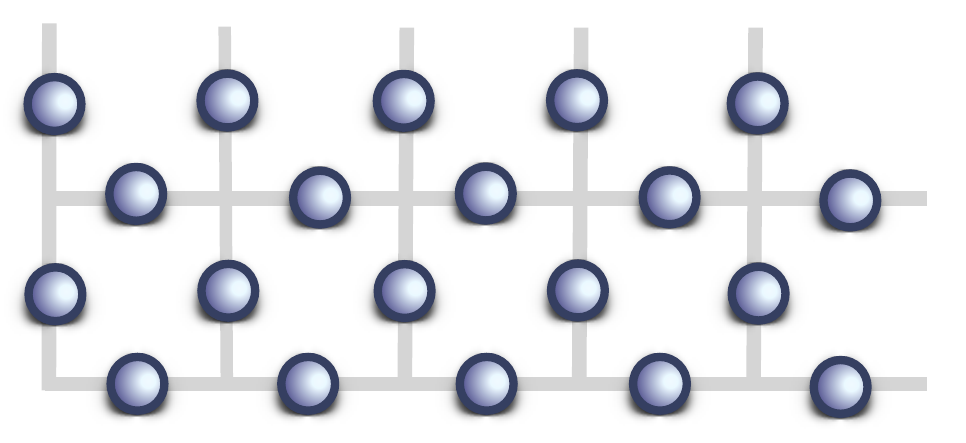} }& \multicolumn{2}{c}{ \includegraphics[width=2cm]{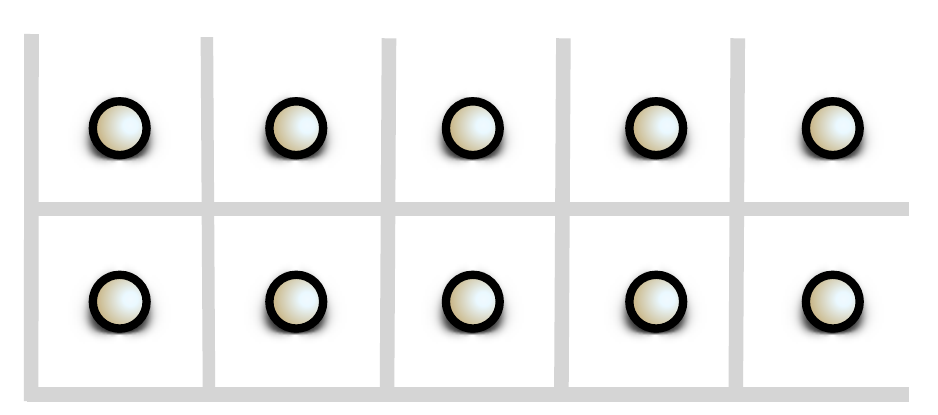} } \\\midrule
        {Magnetic}  & \includegraphics[width=0.8cm]{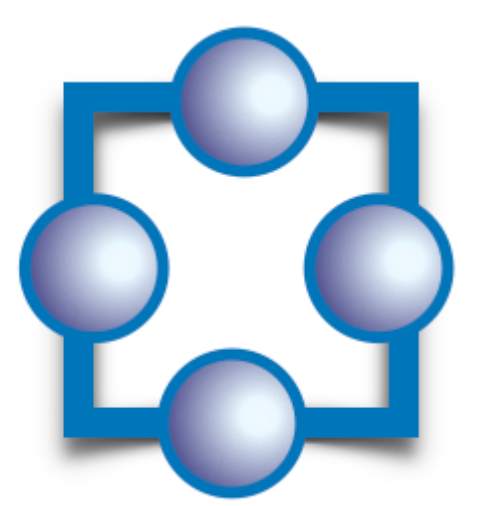}  & $\prod_{\nu \in p} \sigma^x_\nu$ & \includegraphics[width=0.8cm]{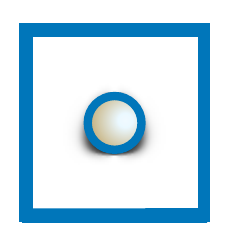}  & $X_p$ \\\midrule
        {Electric}   & \includegraphics[width=0.9cm]{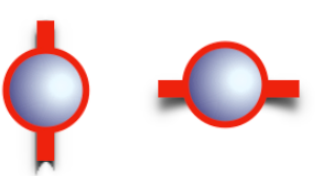} & $\sigma^z_\nu$ &  \includegraphics[width=1.1cm]{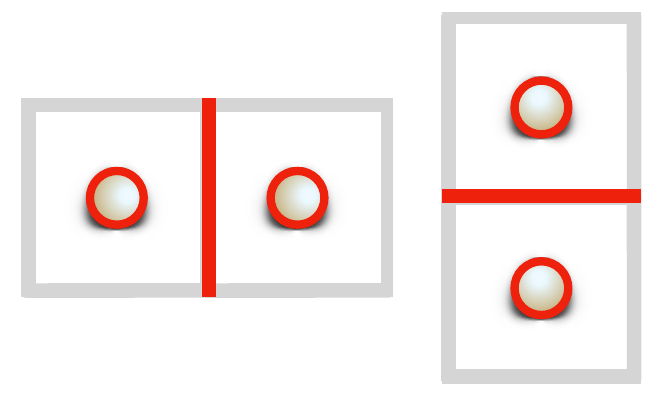} & $Z_{p}Z_{p+\hat{a}}$ \\\midrule
        {Gauss laws}   & \includegraphics[width=0.9cm]{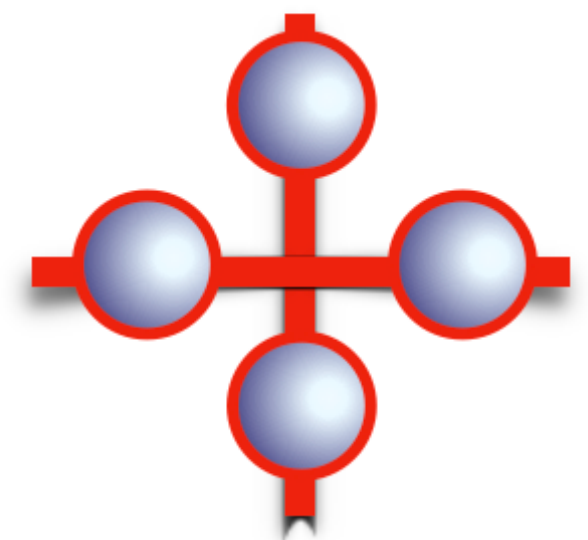}& $\prod_{\nu \in s} \sigma^z_\nu$ & \multicolumn{2}{c}{{Identity}} \\\midrule
        {Parity} & \multicolumn{2}{c|}{{Identity}}  & \multicolumn{2}{c}{$\prod_p X_p=1$}\\ 
        \midrule
         & \includegraphics[width=0.7cm]{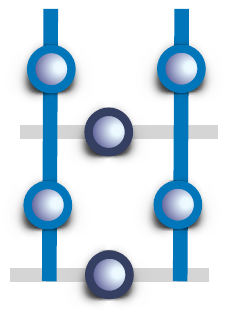}& $\prod_{\nu \in p} \sigma^x_{\nu} \sigma^x_{\nu+L}$ & \includegraphics[width=0.5cm]{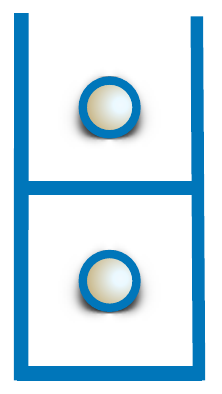} & $X_pX_{p+L}$ \\
         {Stabilizers} & \includegraphics[width=0.7cm]{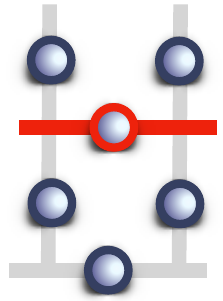}& $\sigma^z_\nu$ & \includegraphics[width=0.5cm]{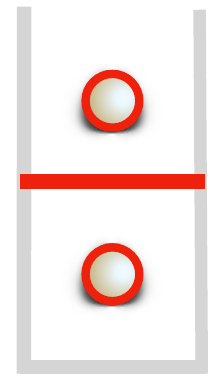} & $Z_pZ_{p+L}$ \\\bottomrule \bottomrule
    \end{tabular}
    \caption{\textit{Mapping between the $\mathbb{Z}_2$ LGT and its Ising dual.} The degrees of freedom are spin-$\frac{1}{2}$ variables, which reside on the links $\nu$ in the LGT and on the faces of the plaquettes $p$ in the dual formulation. The correspondence between magnetic, electric, and Gauss law terms is outlined. In the dual picture, the Gauss laws—originally local constraints in the LGT—is eliminated. The parity constraint in the dual corresponds to the global identity that the product of all plaquettes equals one in the LGT. Additionally, the mapping of the $Z_pZ_{p+L}$ and $X_pX_{p+L}$ stabilizers is presented.\label{tab:LGTdual}}
\end{table}

\subsection{Entanglement Structure and Symmetries}\label{app:entanglement}

\begin{figure}[t]
    \centering
    \includegraphics[width=0.95\linewidth]{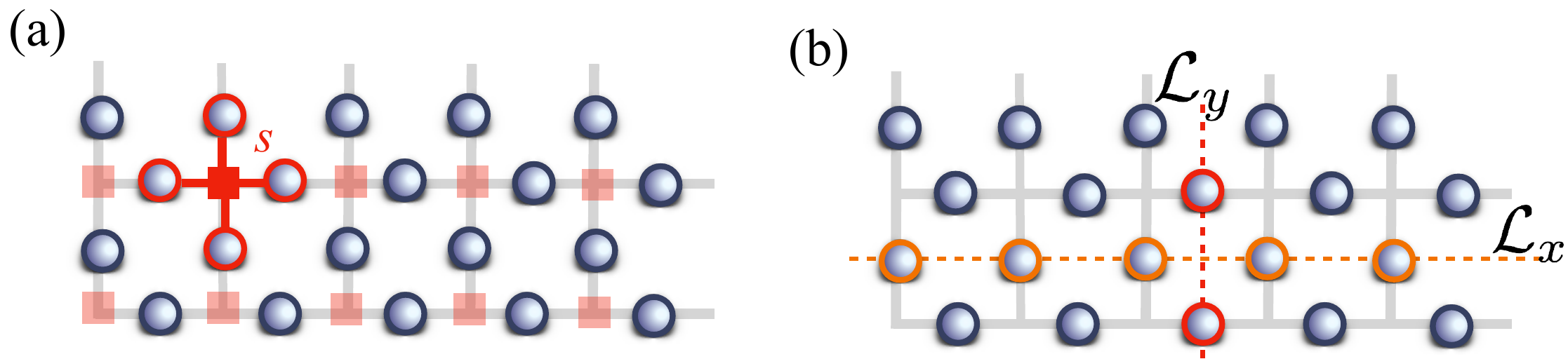}
    \caption{\textit{Symmetries of the LGT.} (a) Shown are $2L$ Gauss laws, \Eq{eq:Glaw}, at sites $s$, which are local symmetry operators that commute with the Hamiltonian $H$, thereby defining superselection sectors. (b) For periodic boundary conditions, the ribbon operators $V_x$ and $V_y$, \Eq{eq:ribop}, defined along $\mathcal{L}_x$ and $\mathcal{L}_y$, respectively, specify additional superselection sectors. \label{fig:symmetriesRDM}}
\end{figure}

In this subsection, we provide additional discussion of the symmetries and entanglement structure of the LGT model and its Ising dual. The (2+1)d $\mathbb{Z}_2$ LGT with periodic boundary conditions exhibits several symmetries, including local Gauss laws, \Eq{eq:Glaw},
at each site or `vertex' $s$, as shown in \Fig{fig:symmetriesRDM}(a). Since $\prod_s G_s = 1$, these operators define $2^{2L-1}$ distinct superselection sectors of the model. We focus on the sector where $G_s = 1$ for all $s$, commonly referred to as the 'physical sector'. Additionally, there are two ribbon operators, depicted in \Fig{fig:symmetriesRDM}(b),
\begin{align}\label{eq:ribop}
    V_{x,y}\equiv \prod_{\nu \in \mathcal{L}_{x/y}}\sigma^z_{\nu}\,,
\end{align}
which wind non-trivially around the two periodic directions of the lattice and commute with the Hamiltonian, i.e., $[H, V_x] = [H, V_y] = 0$. These symmetries result in a total of $2^{2L+1}$ superselection sectors. If a state is an eigenstate of these operators—meaning it is not a superposition of different sectors—then any bipartition into a subsystem $A$ and its complement $\bar{A}$,
\begin{align}
    \rho_A \equiv \text{Tr}_{\bar{A}}(\rho)\,,
\end{align}
will inherit a specific symmetry structure, as discussed in the main text. The remnant of a Gauss operator, shown in \Fig{fig:scars}(b) of the main text, that consists of three of four legs of the Gauss law operator that are inside $A$, at a boundary site $s'$ (i.e. $s'$ is the site just beyond the plaquette $a$ or just before plaquette $b$) is denoted by
\begin{align}\label{eq:symmsec}
   \prod_{\nu \sim s'}\sigma^z_{\nu} = \sigma^z_{\nu^{a/b}_{1/2}}\,,
\end{align}
where ${\nu \sim s'}$ means that only operators within $A$ are included in the product. In the equality, $\nu^{a/b}_{1/2}$ refers to one of  horizontal links just outside of the subsystem, just as shown in \Fig{fig:scars}(b) of the main text, the identity is just a reorganization of Gauss law, $G_s=1$. The argument made in the main text, that any superposition of scar basis states in the LGT for odd $L$ have zero distillable entanglement, is based on $\sigma^z_{\nu'}$, where $\nu'$ is either of the horizontal links $\nu^{a/b}_{1/2}$ in \Fig{fig:scars}(b), and thus $\prod_{\nu \in s'}\sigma^z_{\nu}$ labeling superselection sectors.

To illustrate better that these (four) operators label symmetry sectors, note that they commute with $\rho_A$,
\begin{align}
[ \prod_{\nu \sim s'}\sigma^z_{\nu},\rho_A]=[ \sigma^z_{\nu^{a/b}_{1/2}},\rho_A]=0\,,
\end{align}
and can thus be simultaneously diagonalized, defining symmetry blocks. This can be understood as follows: $\rho_A$ can be expressed as a sum of Pauli strings within $A$,
\begin{align}\label{eq:algebraic}
\rho_A = \frac{1}{2^{d_A}}\sum_{O \in P_A} \langle O \rangle\, O \,,
\end{align}
where $P_A$ are all Pauli strings with non-zero expectation value within $A$, and $d_A$ is the number of spins in $A$. Any of the four operators on the l.h.s of \Eq{eq:symmsec} may be part of this sum and may have a non-zero expectation value. However, no other operator in $A$ fails to commute with it because it is identical, by means of Gauss law, to the r.h.s of \Eq{eq:symmsec}, an operator clearly outside of $A$. Thus $\rho_A$ commutes with it, marking superselection sectors.

The same applies to the ribbon operator shown in \Fig{fig:scars}(c) of the main text, defined as $ V_x^A = \prod_{\nu \sim \mathcal{L}_x^A} $, where $ \mathcal{L}_x^A = \mathcal{L}_x \cap A $. Since $ V_x = V_x^A V_x^{\bar{A}} $, with $ V_x^{\bar{A}} $ being the analogous ribbon operator in the complement of $ A $, the fixed value of $ V_x $ implies, by a similar argument, that $ V_x^A $ defines a symmetry sector of $ \rho_A $. For the other ribbon $V_y$, note that, since the lattice is bipartitioned along $y$, it remains a symmetry operator for both the global state and $\rho_A$; however, as it equals the product of two Gauss laws along $y$, it does not define an independent symmetry block. Although there are six symmetry operators in total (four Gauss laws and two ribbons), only three of them are independent, leading to $2^3=8$ non-zero symmetry blocks.

The entanglement structure of the dual Ising model is fundamentally different from that of the LGT, even though both predict identical expectation values for any operator when mapped between the two formulations. The dual formulation possesses only one symmetry: parity,
\begin{align}
    P_X = \prod_{p=1}^{2L}X_p =1
\end{align}
which corresponds to the requirement that the product of all plaquettes equals one in the LGT, a condition necessary for the LGT-to-Ising mapping under periodic boundary conditions.  This induces a two-block symmetry structure in the reduced density matrix describing any bipartition in the Ising dual. As with the LGT, the parity of $A$, $P_X^A = \prod_{p \in A} X_p$ is a symmetry of $\rho_A$, thereby defining the symmetry blocks: To maintain  $P_X = 1$ parity globally, the subsystem can have either even or odd $X$-parity, or be in any superposition thereof. 

Given the specific entanglement structure arising from the symmetries  of a subsystem in either formulation, it is natural to decompose the von Neumann entanglement entropy into two components: the `symmetry' component, $S^{\rm symm.}$, and the `distillable' component, $S^{\rm dist.}$, as defined in \Eq{eq:vonNeumanncomponents} of the main text. Given the number of symmetry sectors, the symmetry entanglement can be at most $\log(2)$ in the Ising formulation, and $3\log(2)$ in the LGT.
As discussed in the main text and further elaborated below, the scar states in the LGT exhibit zero distillable entanglement (for even and odd $L$) but possess non-zero symmetry entanglement. This characteristic is important: without separating the von Neumann entropy into these components, distinguishing scar states from non-scar states and revealing their stabilizer origin would have been impossible.

\begin{figure}[t]
    \centering
    \includegraphics[width=0.5\linewidth]{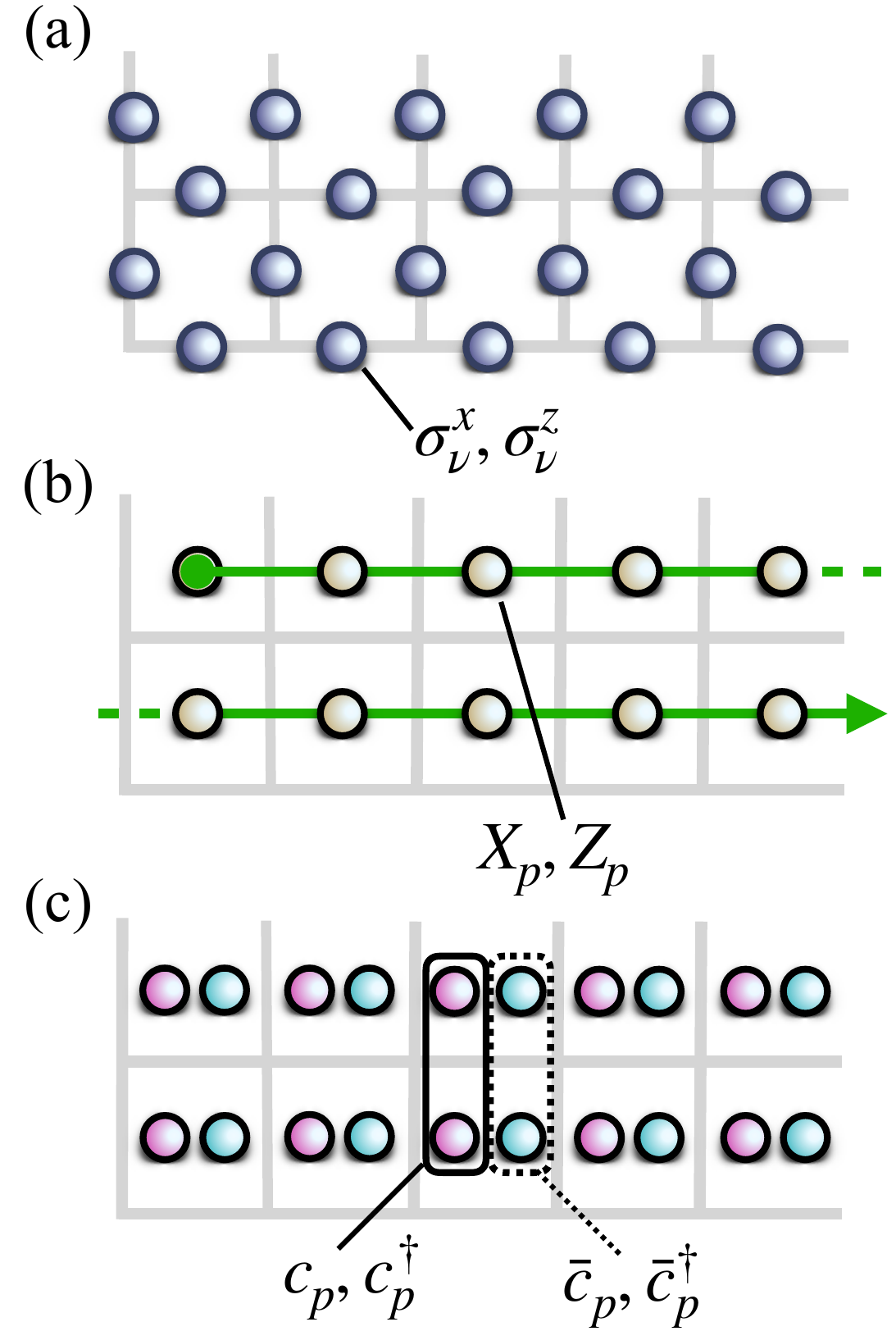}
    \caption{\textit{Analytic Derivation.} (a) The derivation of the scar eigenstates presented in this Letter begins with the degrees of freedom in the $ \mathbb{Z}_2 $ lattice gauge theory (LGT), represented by spins located on the links of a $ L\times 2 $-site lattice with periodic boundary conditions. (b) This system is then mapped onto a dual Ising model, subject to the parity constraint $ \prod_p X_p = 1 $, where spin-${1}/{2} $ degrees of freedom reside at the centers of the plaquettes (denoted by $ p = 1, \ldots, 2L $). A Jordan-Wigner transformation (indicated by the green arrow) initiates in the top row (covering plaquettes $ 1, \ldots, L $) and extends to the bottom row (from sites $ L+1 $ to $ 2L $), transforming the system into a two-component Majorana representation. (c) The Majorana modes are combined vertically to form a two-flavor fermion representation, where $ p = 1, \ldots, L $ denotes one-dimensional sites. This representation enables the identification of scar subspaces for both even and odd $ L $ via  the commutant  of the algebra generated by the terms in $H$.
    \label{fig:transformations}}
\end{figure}

\subsection{Derivation of the Scar Solutions}\label{app:derivation}
The $\mathbb{Z}_2$ LGT model with periodic boundary conditions, \Eq{eq:Z2Hamiltonian} of the main text, can be mapped onto a two-leg transverse field Ising model (\TFI),\Eq{eq:IsingDualHamiltonian} of the main text, under the constraint $\prod_pX_p=1$ which ensures that the product of all plaquettes is one, an identity on the LGT side, see \Fig{fig:transformations}(a) and (b). The \TFI~   is mapped onto a fermionic system via a Jordan-Wigner transformation that runs from the first spin in the top row to the last spin in the top row, then continuing at the first spin on the bottom row and ending at the last spin in the bottom row, see \Fig{fig:transformations}(b). As shown in \Fig{fig:transformations}(c), this is written in terms of Majorana operators, $\gamma_p,\bar{\gamma}_{p}$, $p=1,\dots,2L$ with $\gamma_p^2=\bar{\gamma}_p^2=1$ and $\{ \gamma_p,\gamma_q\} = \{ \bar{\gamma}_p,\bar{\gamma}_q\}= 2\delta_{pq}$,  $\{ \gamma_p,\bar{\gamma}_q\} =0$, as follows
\begin{align}\label{eq:IsingtoMajorana}
    X_p = -i \gamma_p \bar{\gamma}_p\;, \quad Z_p = \prod_{q=1}^{p-1}(i\gamma_q \bar{\gamma}_q)\gamma_p\,.
 \end{align}
 Next, we define two species fermionic operators by combining the Majorana modes \textit{vertically along $y$}, i.e.,
 \begin{align}\label{eq:majortoferm}
    &c_p=\frac{1}{2}(\gamma_p - i\gamma_{p+L})\,,\quad \bar{c}_p=\frac{1}{2}(\bar{\gamma}_p - i\bar{\gamma}_{p+L})\,,\nonumber\\
        &c_p^\dagger=\frac{1}{2}(\gamma_p + i\gamma_{p+L})\,,\quad \bar{c}_p^\dagger=\frac{1}{2}(\bar{\gamma}_p + i\bar{\gamma}_{p+L})\,.
\end{align}
The fermion parity operators are given as
 \begin{align}
(-1)^{c_p^\dagger c_p}=i\gamma_p \gamma_{p+L}\,,\quad (-1)^{\bar{c}_p^\dagger \bar{c}_p}=i\bar{\gamma}_p \bar{\gamma}_{p+L}\,,
\end{align}
so that 
\begin{align}\label{eq:locpar}
    X_pX_{p+L} \equiv -(-1)^{c_p^\dagger c_p+\bar{c}_p^\dagger \bar{c}_p}\,,
\end{align}
marks the combined-species fermion parity. We define the vacuum state by requiring $ c_p| 0 \rangle =\bar{c}_p| 0 \rangle=0$ for all $p$, with the mode ordering from left to right in ascending order, i.e., $c_1^\dagger \bar{c}_1^\dagger \dots  c_L^\dagger \bar{c}_L^\dagger| 0\rangle$. Note that the constraint
\begin{align}
1=\prod_{p=1}^{2L} X_p =\prod_{p=1}^{L} X_pX_{p+L}=(-1)^L\prod_{p=1}^{L}(-1)^{c_p^\dagger {c}_p+\bar{c}_p^\dagger\bar{c}_p }\,,
\end{align}
either represents the fermion parity for even $L$ or the \textit{negative} fermion parity for odd $L$. Consequently, for even $L$, only an even number of fermion or hole excitations are allowed, whereas for odd $L$, only an odd number of fermion or hole excitations are possible. This property is crucial for the scar solutions.

In this representation, two of the three terms of the Ising Hamiltonian, Eq. (2) of the main text, conserve the total fermion number,
\begin{align}\label{eq:particlenumber}
    \hat{N}=\sum_{p=1}^L(c_p^\dagger c_p +\bar{c}_p^\dagger \bar{c}_p)\,,
\end{align}
which ranges from $\hat{N}=0$ to $\hat{N}=2L$.
Concretely, using \Eqs{eq:IsingtoMajorana}{eq:majortoferm}, we can rewrite  the first  term in \Eq{eq:IsingDualHamiltonian} of the main text as
\begin{align}
    X_p + X_{p+L} &= i\bar{\gamma}_p \gamma_p + i\bar{\gamma}_{p+L} \gamma_{p+L}\nonumber\\
    &=2i(\bar{c}_p {c}^\dagger_p + \bar{c}_p^\dagger {c}_p )\,.
\end{align}
A similar derivation for the third term, for $p<L$, yields 
\begin{align}
    Z_pZ_{p+1}+Z_{p+L}Z_{p+L+1} &= i\bar{\gamma}_p\gamma_{p+1} +  i\bar{\gamma}_{p+L}\gamma_{p+L+1}\nonumber\\
    &= 2i(\bar{c}_p c^\dagger_{p+1} + \bar{c}^\dagger_p c_{p+1})\,,
\end{align}
demonstrating that both terms conserve particle number. For $p=L$, however, we find
\begin{align}\label{eq:term1}
    &Z_1 Z_L + Z_{L+1} Z_{2L}
= \prod_{q=1}^L (i\gamma_q \bar{\gamma}_q)\, (i \gamma_1 \bar{\gamma}_L+i \gamma_{L+1} \bar{\gamma}_{2L})
\end{align}
where $i \gamma_1 \bar{\gamma}_L+i \gamma_{L+1} \bar{\gamma}_{2L}=2i (c_1^\dagger \bar{c}_L + c_1 \bar{c}_L^\dagger)$ conserves particle number, but the string $\prod_{q=1}^L (i\gamma_q \bar{\gamma}_q)$  does not. This string arises because $Z_1 Z_L + Z_{L+1} Z_{2L}$ is a non-local operator connecting across the periodic boundary condition, reflecting the choice of Jordan-Wigner map between spins and fermions shown in \Fig{fig:transformations}(b). Similarly, the remaining term in \Eq{eq:IsingDualHamiltonian}  of the main text also involves also a Jordan-Wigner string:
\begin{align}\label{eq:term2}
    Z_p Z_{p+L} = i \prod_{q=0}^{L-1}(i\gamma_{p+q} \bar{\gamma}_{p+q} ) \, i\gamma_p \gamma_{p+L}
\end{align}
which does not conserve fermion number either.

Despite not conserving particle number, both \Eq{eq:term1} and \Eq{eq:term2} exhibit particle-hole symmetry, i.e., conjugating $\hat{N}$ to $2L-\hat{N}$ and vice versa. Consequently, all terms in the Hamiltonian conserve $(\hat{N}-L)^2$. As we will show below, the scar solutions are to be found, for even $L$, in the $(\hat{N}-L)^2=L^2$ subspace, corresponding to the completely empty, $\hat{N}=0$, or fully filled, $\hat{N}=2L$, states. 
For odd $L$, the scars lie in the $(\hat{N}-L)^2 = (1-L)^2= (2L-1-L)^2$ subspace, i.e., the combined single-particle, $\hat{N}=1$, and single-hole, $\hat{N}=2L-1$, states.

Equipped with this, we will now explicitly demonstrate that Eq. (2) of the main text posses exact scar eigenstates for even $L$ which are the Fock vacuum and the fully occupied states, and furthermore that for odd $L$ the scar subspace is spanned by $4L$ single-particle and single-hole excitations of the $\bar{c}^\dagger$ and ${c}^\dagger$ modes, $ H^{\rm dual}$ has no matrix elements connecting these to any other state.

\subsubsection{Scar Solutions for even $L$}
Given the properties discussed in the previous section, we posit that the two scar solutions for even $L$ are found in the smallest eigenspaces of $(\hat{N}-L)^2$: the vacuum state $| 0 \rangle$ and the fully occupied state 
\begin{align}\label{eq:fullyocc}
| \Lambda \rangle \equiv \prod_p c_p^\dagger\bar{c}_p^\dagger | 0 \rangle 
\end{align}
where the product is in ascending order. These are eigenstates with eigenvalue zero of all terms in \Eq{eq:IsingDualHamiltonian}  of the main text, and by extension of \Eq{eq:Z2Hamiltonian}  of the main text,  which we will show now. Both $| 0 \rangle $ and $| \Lambda \rangle$ are uniquely specified via the fermion parity, specifically for these states
\begin{align}\label{eq:Cond1even}
  1=  (-1)^{c_p^\dagger c_p + \bar{c}^\dagger_p \bar{c}_p} = -X_p X_{p+L}\,,
\end{align}
where we used \Eq{eq:locpar}, and 
\begin{align}\label{eq:Cond2even}
1&=(-1)^{\bar{c}^\dagger_p \bar{c}_p + c_{p+1}^\dagger c_{p+1}} = \bar{\gamma}_p\gamma_{p+1}\bar{\gamma}_{p+L}\gamma_{p+L+1}
\nonumber\\
&=-Z_{p}Z_{p+1} Z_{p+L}Z_{p+L+1}\,,
\end{align}
where in the last equality we have used that $\gamma_p\equiv \prod_{k=1}^{p-1}(-X_k)Z_p$ and $\bar{\gamma}_p\equiv -iX_p \prod_{k=1}^{p-1}(-X_k) Z_p$. From \Eq{eq:Cond1even} it follows that the space spanned by  $| 0 \rangle $ and $| \Lambda \rangle$ must have $X_p X_{p+L} =-1$ and from \Eq{eq:Cond2even} that $Z_p Z_{p+L} = -Z_{p+1} Z_{p+L+1} $ for all $p$. We can thus pick the two states given in \Eq{eq:scarevenL1} of the main text to span the even-$L$ subspace. 

We will show now that these are \textit{exact} eigenstates with eigenvalue zero for any value of the coupling $g$. Specifically, $\rho_{1/2}$ are eigenstates of, i.e. commutes with, every individual term in \Eq{eq:scarevenL1} of the main text. We write $H^{\rm dual}\equiv H_A + H_B +H_C$ where
\begin{align}\label{eq:HdualA}
        H_A\equiv  &- \sum_{p=1}^L [X_p+X_{p+L}]\,,\\\label{eq:HdualB}
        H_B\equiv  & - \tilde{g}\sum_{p=1}^L Z_pZ_{p+L} \,,\\
     H_C \equiv &-\sum_{p=1}^L g_p [Z_p Z_{p+1} + Z_{p+L} Z_{p+L+1} ]\,.\label{eq:HdualC}
\end{align}
First note that if $X_p X_{p+L} =-1$ (\Eq{eq:Cond1even}), then $X_p = - X_{p+L}$, and therefore  $H_A | \Psi _{1/2}\rangle =0 $ (or alternatively $[H_A,\rho_{1/2}]=0$). Furthermore, from \Eq{eq:Cond2even} it also follows that
\begin{align}
Z_pZ_{p+1} = -Z_{p+L}Z_{p+L+1}\,
\end{align}
and thus  $H_C | \Psi _{1/2}\rangle =0 $. Finally, note that
\begin{align}
    H_B | \psi_{1/2} \rangle = -\tilde{g}  \sum_{p=1}^{L} \pm(-1)^p | \psi_{1/2} \rangle=0\,,
\end{align}
because $L$ is even. Therefore $|\psi_{1/2}  \rangle$ are degenerate exact eigenstates of $H^{\rm dual}$ with eigenvalue zero (and thus of \Eq{eq:Z2Hamiltonian} of the main text on the other side of the duality). Because the states $| \psi_{1/2}\rangle$, as well as $| 0 \rangle$ and $| \Lambda \rangle$, are product states, their von Neumann entanglement entropy is zero for any bipartition. More specifically, separate symmetry and distillable entanglement components can be defined via \Eq{eq:vonNeumanncomponents} of the main text. Both the distillable and the symmetry component are zero, $S_{\rm dist. }^{\rm dual}=S_{\rm sym }^{\rm dual}=0$, for the even-$L$ scars in the Ising formulation. In the LGT formulation the distillable entanglement is also zero, $S_{\rm dist. }^{\rm LGT}=0$, but the scars are not product states. A non-zero symmetry component stems from the Gauss laws at the boundary separating a subsystem from its complement, see the previous discussion in Section "Entanglement Structure and Symmetries" of this Supplemental Material. 

\subsubsection{Scar Subspace for odd $L$}
For odd $L$, we now discuss a subspace spanned by scar states. The states $| 0 \rangle$ and $| \Lambda\rangle$ are not allowed for odd $L$ because of \Eq{eq:locpar}, instead only states with an odd fermion-parity are permitted. We posit that the scar subspace is spanned by the following $4L$ single-particle and single-hole states,
\begin{align}
    | \psi_p \rangle &\equiv c_p^\dagger | 0 \rangle\,,\quad | \bar{\psi}_p \rangle \equiv \bar{c}_p^\dagger | 0 \rangle\,,\label{eq:particlestates}\\
    | \chi_p \rangle &\equiv c_p | \Lambda  \rangle\,,\quad | \chi_p \rangle \equiv -\bar{c}_p | \Lambda \rangle\,,\label{eq:holestates}
\end{align}
$| \Lambda \rangle$ is defined in \Eq{eq:fullyocc} and the fermion-parity is consistent with \Eq{eq:locpar} for odd $L$. The terms in the Hamiltonian can be written in terms of Majorana operators and, subsequently, via fermion Fock operators. The plaquette term is
\begin{align}\label{eq:majoranaH1}
H_A= -\sum_{p=1}^L(i\bar{\gamma}_p \gamma_{p} +i\bar{\gamma}_{p+L} \gamma_{p+L}  )
=-2i \sum_{p=1}^L(\bar{c}_p^\dagger c_p - c_p^\dagger \bar{c}_p)\,,
 \end{align}
which conserves the total fermion number, \Eq{eq:particlenumber}.
The electric terms in the Hamiltonian can be written as
  \begin{align}\label{eq:majoranaH2}
H_B = \tilde{g}\sum_{p=1}^L\prod_{q=0}^{L-1}(i\gamma_{p+q} \bar{\gamma}_{p+q})\, \gamma_p\gamma_{p+L}
   \end{align}
 and
 \begin{align}\label{eq:majoranaH3}
     H_C=&-g \sum_{p=1}^{L-1}(i\bar{\gamma}_p \gamma_{p+1}+i\bar{\gamma}_{p+L} \gamma_{p+L+1})
   \nonumber  \\&-gV_x\prod_{q=1}^L(i\gamma_q \bar{\gamma}_q) (i\gamma_1 \bar{\gamma}_L+i\gamma_{L+1} \bar{\gamma}_{2L}) 
 \end{align}
 whose properties are more difficult to discern. To understand the action of $H_C$ (\Eq{eq:majoranaH3}) note that
 \begin{align}
     i\bar{\gamma}_p \gamma_{p+1}+i\bar{\gamma}_p \gamma_{p+1} &= 2i(\bar{c}_p^\dagger c_{p+1} -c_{p+1}^\dagger \bar{c}_p )\,,
     \\
     i\gamma_1 \bar{\gamma}_L+i\gamma_{L+1} \bar{\gamma}_{2L} &= 2i(c_1^\dagger \bar{c}_L-\bar{c}_p^\dagger c_1)\,,
 \end{align}
 are particle number conserving terms. Importantly, the product in the second line of \Eq{eq:majoranaH3} can be written as
\begin{align}
    \prod_{q=1}^L(i\gamma_q \bar{\gamma}_q) =  \prod_{q=1}^L i(c^\dagger_q \bar{c}^\dagger_q+c^\dagger_q \bar{c}_q+c_q \bar{c}^\dagger_q+c_q \bar{c}_q)
\end{align}
and it conserves the space spanned by \Eqs{eq:particlestates}{eq:holestates}. Specifically, one can show that
\begin{align}
     &\prod_{q=1}^L(i\gamma_q \bar{\gamma}_q) | \psi_p\rangle = -i^L | \chi_p\rangle\,,\nonumber
     \\
     &\prod_{q=1}^L(i\gamma_q \bar{\gamma}_q) | \bar{\psi}_p\rangle = i^L | \bar{\chi}_p\rangle\,,\nonumber
     \\
     &\prod_{q=1}^L(i\gamma_q \bar{\gamma}_q) | \chi_p\rangle = i^L | \psi_p\rangle\,,\nonumber
     \\
     &\prod_{q=1}^L(i\gamma_q \bar{\gamma}_q) | \bar{\chi}_p\rangle = -i^L | \bar{\psi}_p\rangle\,.
\end{align}
Moreover, for \Eq{eq:majoranaH2}, one finds that $\gamma_p \gamma_{p+L}=2i c_p^\dagger c_p$ and that
\begin{align}
&\prod_{q=0}^{L-1}(i\gamma_{p+q} \bar{\gamma}_{p+q}) | \psi_p\rangle = \,i^L(-1)^p \theta({k,q})|\chi_p\rangle\,,\nonumber
     \\
  &   \prod_{q=0}^{L-1}(i\gamma_{p+q} \bar{\gamma}_{p+q}) | \bar{\psi}_k\rangle = -i^L(-1)^p \theta({k,q})|\bar{\chi}_k\rangle\,,\nonumber
     \\
   &  \prod_{q=0}^{L-1}(i\gamma_{p+q} \bar{\gamma}_{p+q}) | \chi_k\rangle = -i^L(-1)^p \theta({k,q})|\psi_k\rangle\,,\nonumber
     \\
   &  \prod_{q=0}^{L-1}(i\gamma_{p+q} \bar{\gamma}_{p+q}) | \bar{\chi}_k\rangle = i^L(-1)^p \theta({k,q})|\bar{\psi}_k\rangle\,,
\end{align}
where $\theta({k,q})\equiv1$ if $k\le q$ and $-1$ otherwise. This establishes that $H^{\rm dual}$ (and thus $H$) has no matrix elements between the scar and non-scar subspaces.
\begin{figure*}[t]
    \centering
    \includegraphics[width=0.8\linewidth]{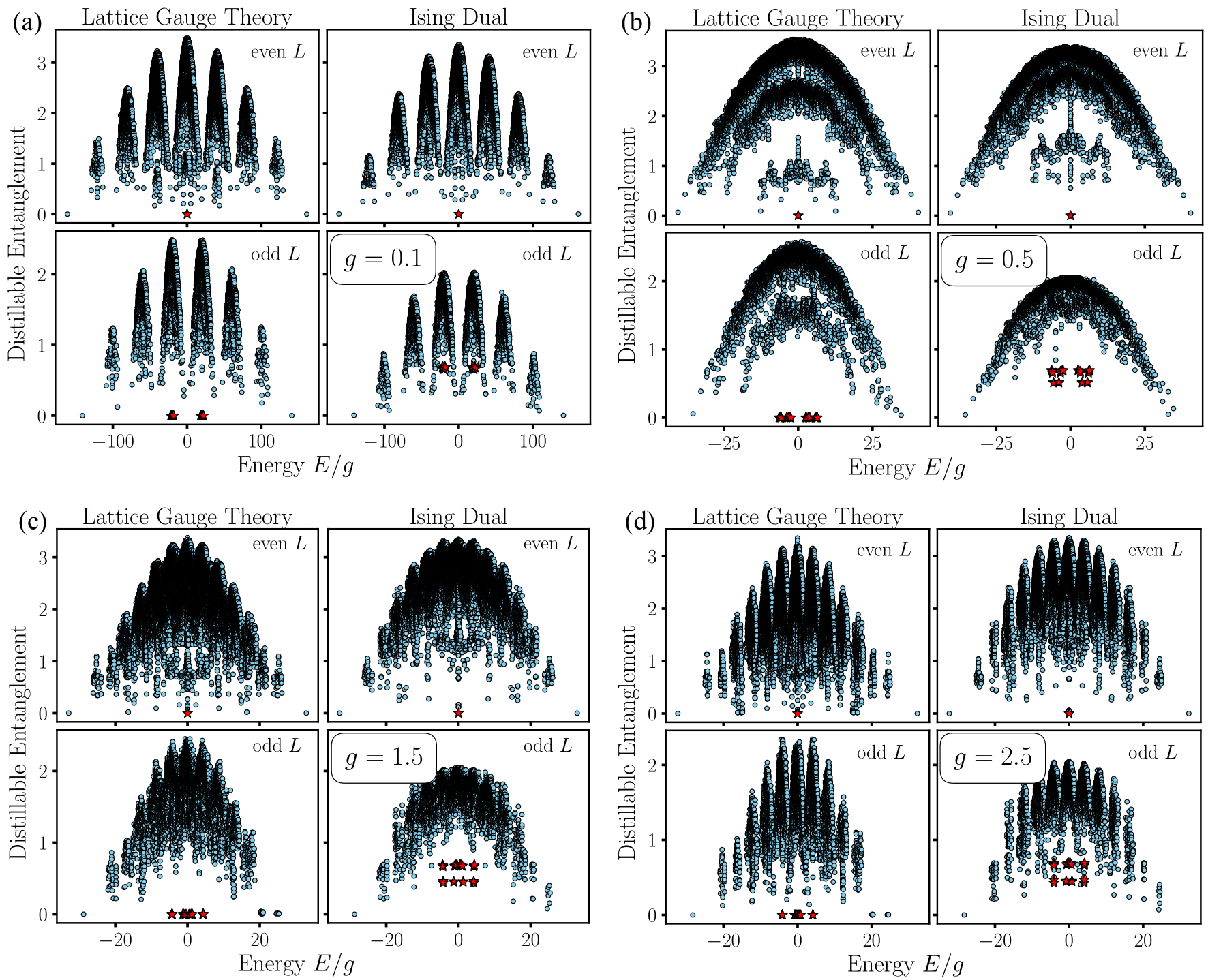}
    \caption{\textit{Distillable Entanglement for varying Couplings.} (a) The plot shows $S^{\rm dist}_{\rm vN} $ for $g = 0.1 $, close to the integrable point $g = 0$. Panels in the top row display results for an even system size $ L = 8 $, while the bottom row shows results for an odd system size $L = 7$. The left panels correspond to results from the LGT formulation, and the right panels to those from the Ising model. Scar states are highlighted with red stars. In the Ising dual formulation, scar states exhibit non-zero distillable entanglement, blending with other typical states and becoming indistinguishable. However, in the LGT, scar states maintain exactly zero distillable entanglement, which differentiates them clearly. (b) Results for $ g = 0.5 $. We point out a band structure visible within the typical states. This structure may be due to other (exact or approximate) protected subspaces in the model. (c) $ g = 1.5$. (d)  $ g = 2.5$. While the limits $ g \rightarrow 0$ and $g \rightarrow \infty $ yield ground states that are straightforward to understand, the behavior of mid-spectrum states is complex, with scar eigenstates persisting in these limits.}
    \label{fig:distill}
\end{figure*}%

While a rather lengthy derivation brought these results, a more elegant solution can be found noting that \Eqs{eq:particlestates}{eq:holestates} are uniquely specified by
\begin{align}\label{eq:oddLCond1}
    (-1)^{c_k^\dagger c_k + \bar{c}^\dagger_k \bar{c}_k} =  \begin{cases}
        1 & \text{if } k= p\\
        -1 & \text{otherwise}  
    \end{cases}\,,
\end{align}
specifying the values of $X_p X_{p+L}$, and 
\begin{align}\label{eq:oddLCond2}
(-1)^{\bar{c}^\dagger_p \bar{c}_p + c_{p+1}^\dagger c_{p+1}} 
=\begin{cases}
        -1 & \text{if } k= p \text{ or } k=p-1 
        \\ 1 & \text{otherwise}
        \end{cases}
\end{align}%
\begin{figure*}[t]
    \centering
    \includegraphics[width=0.8\linewidth]{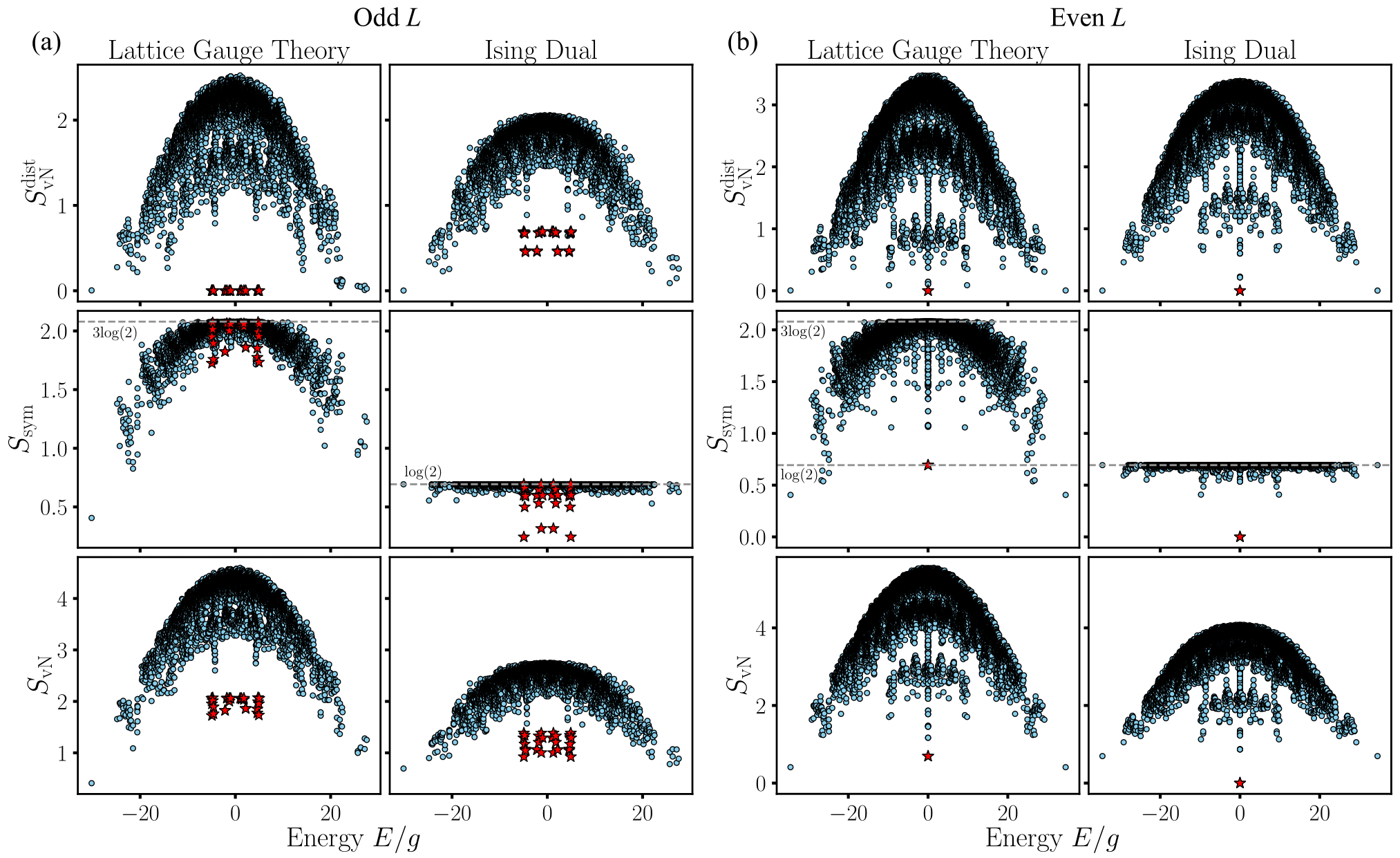}
    \caption{\textit{Contributions to von Neumann Entanglement Entropy.} The top row shows the distillable entanglement $ S^{\rm dist}_{\rm vN} $, the middle row the symmetry entanglement $ S^{\rm symm}_{\rm vN} \equiv - \sum_s p_s \log(p_s) $, and the bottom row the total von Neumann entropy $ S_{\rm vN} = S^{\rm dist}_{\rm vN} + S^{\rm symm}_{\rm vN} $. The left column contains results from the LGT, while the right column shows results for the Ising dual. Quantum many-body scar (QMBS) states are highlighted in red, with the maximum possible symmetry entanglement indicated by horizontal gray dashed lines. (a) Results for  odd system size $ L = 7 $ and (b)  for  even system size $ L = 8 $. For $ g = 0.9 $, these results show that QMBS states can be clearly distinguished by their distillable entanglement in the LGT, where they exhibit exactly zero values. By contrast, the total von Neumann entropy, although widely used, is finite and therefore less reliable for identifying QMBS states. Indeed, at other coupling values and in smaller systems, scar states may blend with typical states in terms of $ S_{\rm vN} $ even within the LGT, rendering them indistinguishable. } 
    \label{fig:contrib}
\end{figure*}%
specifying the values of $Z_p Z_{p+L}$. The $4L$ states obeying \Eqs{eq:oddLCond1}{eq:oddLCond2} are given in \Eq{eq:scaroddL1} of the main text. We will now work out the matrix elements of the Hamiltonian in these states. To do so, we first write them as
\begin{align}\label{eq:oddLscarsolutionsstates}
 |\varphi_\alpha(k) \rangle\equiv& 
\prod_{q\neq k}\Big[\frac{1-X_q X_{q+L}}{\sqrt{2}} \Big] \frac{1+X_kX_{k+L}}{\sqrt{2}} \, | r_\alpha(k)\rangle 
\end{align}
where $| r_\alpha(k) \rangle $
are $Z$-eigenstates that realize the desired $Z_p Z_{p+L}$ stabilizer property (but are not $X_pX_{p+L}$ eigenstates). Choosing a convention, where the spin in the bottom row is always up, the spin in the top row selects the $Z_p Z_{p+L}$ eigenvalue, for instance for $\alpha=1$,
\begin{align}
    |r_1(k) \rangle \equiv \prod_{p<k} \left| 
\genfrac{}{}{0pt}{}{ s^{(p,k)}}{\ua } \right\rangle  
\; \overbrace{\left|\genfrac{}{}{0pt}{}{ \da}{\ua } \right\rangle}^{k}\;
\prod_{p>k} \left| 
\genfrac{}{}{0pt}{}{ \bar{s}^{(p,k)} }{\ua } 
\right\rangle\,,
\end{align}
where $s^{(p,k)}=\ua$ if $k-q$ is even and otherwise $\da$, and $\bar{s}^{(p,k)}=\da$ if $k-1$ is even and otherwise $\ua$. The cases $\alpha=2,3,4$ follow directly from \Eq{eq:scaroddL1} of the main text. The matrix elements of $H^{\rm dual}$ \Eq{eq:IsingDualHamiltonian} of the main text with individual terms given in \Eqs{eq:HdualA}{eq:HdualA}), and identically for the LGT Hamiltonian Eq. (1) of the main text, are now easily derived. For the plaquette term, $ H_{A}=\sum_{p}(X_p+X_{p+L})$, the scar-subspace matrix elements read
\begin{align}
    \langle \varphi_4(k) | H_{A} | \varphi_1(k) \rangle = \langle \varphi_1(k) | H_{A} | \varphi_4(k) \rangle &\nonumber\\
   = \langle \varphi_3(k) | H_{A} | \varphi_2(k) \rangle = \langle \varphi_2(k) | H_{A} | \varphi_3(k) \rangle &= 2\,,
\end{align}
while the scar-subspace matrix elements of $ H_{B}=\sum_{p}Z_pZ_{p+L}$ are 
\begin{align}
\langle \varphi_1(k) | H_{B}| \varphi_1(k) \rangle =\langle \varphi_3(k) | H_{B}| \varphi_3(k) \rangle &=-1\,,\nonumber\\
\langle \varphi_2(k) | H_{B}| \varphi_2(k) \rangle =\langle \varphi_4(k) | H_{B}| \varphi_4(k) \rangle &=+1\,.
\end{align}
Finally, the matrix elements of $ H_{C}=\sum_p(Z_{p}Z_{p+1}+Z_{p+L}Z_{p+1+L})$ are
\begin{align}
  \langle \varphi_3(k-1)  |H_{C}| \varphi_1(k) \rangle=\langle \varphi_1(k+1)  |H_{C}| \varphi_3(k) \rangle& \nonumber\\
  =\langle \varphi_4(k-1) |H_{C}| \varphi_2(k) \rangle=\langle \varphi_2(k+1) |H_{C}| \varphi_4(k) \rangle &=2\,.
\end{align}
While states are represented differently in the LGT, see the main text, these matrix elements  are identical between the formulations.

\subsection{Details of the Quantum Many-body Scar Solutions}\label{app:detailsQMBS}
In this section of the Supplemental Material we provide additional numerical details for the QMBS  solution discussed in the main text, both in the LGT and the Ising formulation. All results are for $V_x=V_y=1$, though while the $V_y=-1$ sector is trivial, the QMBS solutions exist in all sectors.

\begin{figure*} 
    \centering
    \includegraphics[width=0.8\linewidth]{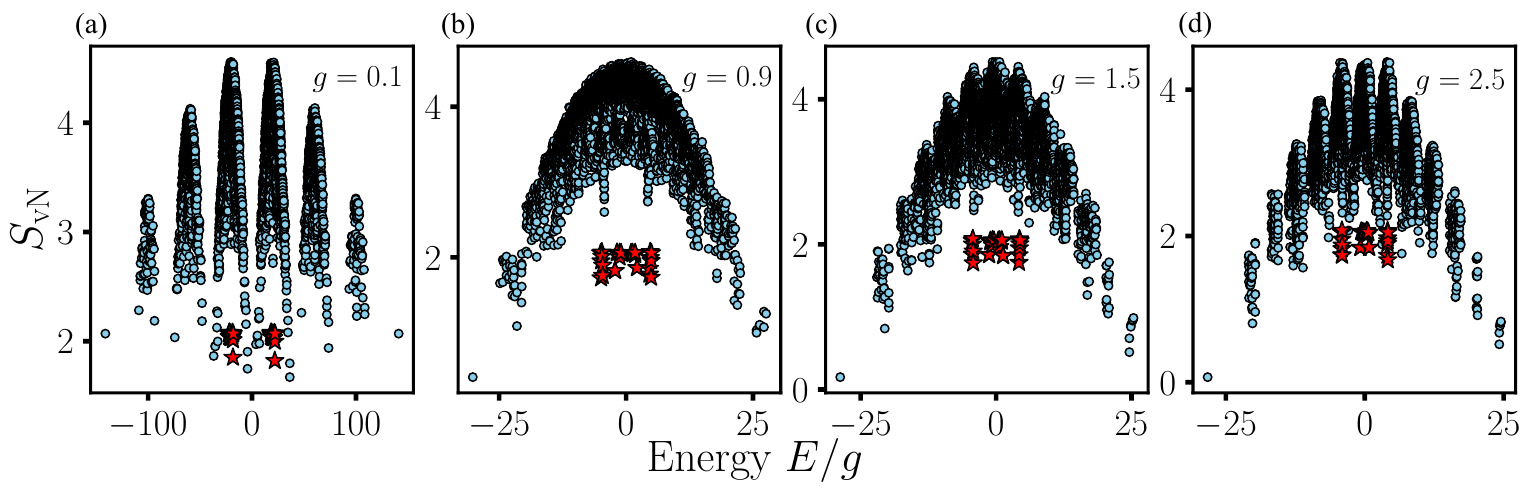}
    \caption{\textit{Total von Neumann Entanglement Entropy.} Shown is the (total) Neumann Entropy $S_{vN}$ for lattice size $L=7$ with different couplings and with the scars, identified via their vanishing \textit{distillable} entanglement, highlighted in red. (a) $g=0.1$. (b) $g=0.9$, (c) $g=1.5$, (d) $g=2.5$, Especially at high and low coupling, panels (a) and (d), the spectrum starts separating into energy bands and identifying the scars becomes impossible based on $S_{\rm vN}$ when not separated into distillable and symmetry components. For smaller sized systems, distinguishing scars from non-scars based on $S_{vN}$, instead of via the {distillable} entanglement, becomes impossible even at intermediate couplings.}
    \label{fig:total_VN}
\end{figure*}

\Fig{fig:distill} shows the distillable entanglement, for even $L=8$ and odd $L=7$, across a range of couplings, $g\in \{0.1,0.5,1.5,2.5\}$. The scar solutions persist for any value of the coupling and, moreover, for the LGT have exactly zero entanglement (which we verified to within machine precision). In contrast, in the Ising dual the distillable entanglement is non-zero and, depending on the coupling, the scar solutions can be indistinguishable from typical eigenstates. Notably, while the limites $g\rightarrow 0$ and $g\rightarrow \infty$ are well understood for the ground states of the model, mid-spectrum states behave highly non-trivial in these limits with the scar states persisting. Additionally, in  \Fig{fig:distill}(b) we observe an intricate band structure within the typical eigenstates. We speculate that these are due to additional protected subspaces potentially those that, in the fermionic formulation, are singled out by $(\hat{N}-L)^2$, i.e., the combined $k$-particle and -hole spaces where $k=2,\dots$, which an extension of our investigation may uncover.

\Fig{fig:contrib} shows the different components of the von Neumann entanglement entropy, the distillable and the symmetry part. Shown are the  distillable entanglement $ S^{\rm dist}_{\rm vN} $ in the top row, the symmetry entanglement $ S^{\rm symm}_{\rm vN} \equiv - \sum_s p_s \log(p_s) $ in the middle row, and the total von Neumann entropy $ S_{\rm vN} = S^{\rm dist}_{\rm vN} + S^{\rm symm}_{\rm vN} $ in the bottom row. The QMBS states are highlighted in red, all results are for $L=7$ (odd) and $L=8$ (even) lattice sizes, with coupling $g=0.9$.  Although the total von Neumann entropy of scar states remains somewhat distinguishable at this coupling, it tends to blend scar states with non-scar states at other couplings, potentially obscuring the scar solutions. This underscores the value of studying entanglement structure—rather than only entanglement entropy — as a more revealing marker to differentiate scar from non-scar states. Our analytical analysis shows that, because of their stabilizer origin, it is magic resource that truly differentiates scar from non-scar states.

To emphasize this point more, Fig.~\ref{fig:total_VN} shows the (total) von Neumann entropy $ S_{\rm vN} $ for $ L = 7 $ and various coupling strengths, $ g \in \{0.1, 0.9, 1.5, 2.5\} $. For both small and large couplings, scar and non-scar eigenstates become indistinguishable as the scar states merge into the continuum of non-scar states. This underscores the importance of identifying them through their distillable entanglement, which remains exactly zero.

Finally, we note a technical but important detail in our numerical analysis. For even $L$, the scar states are degenerate in energy. Since we are using exact diagonalization, our numerics typically return a linear superposition of the two scar states, which are analytically identified in the main text, and this superposition often has finite distillable entanglement. To resolve this, we employ a standard technique to break the degeneracy by an infinitesimal amount, ensuring that the scar subspace remains intact while making the two scar solutions infinitesimally non-degenerate. Specifically, we perturb the coupling in front of $Z_p Z_{p+L}$ for a randomly chosen $p$ by modifying the coupling as $g \rightarrow g + \epsilon$, where $\epsilon = 0.001$. In cases where analytic solutions are unavailable, an alternative method to handle this issue is described in~\cite{biswas2022scars}.

\subsection{Other Lattice Geometries}
\begin{figure}[t]
    \centering
    \includegraphics[width=0.75\linewidth]{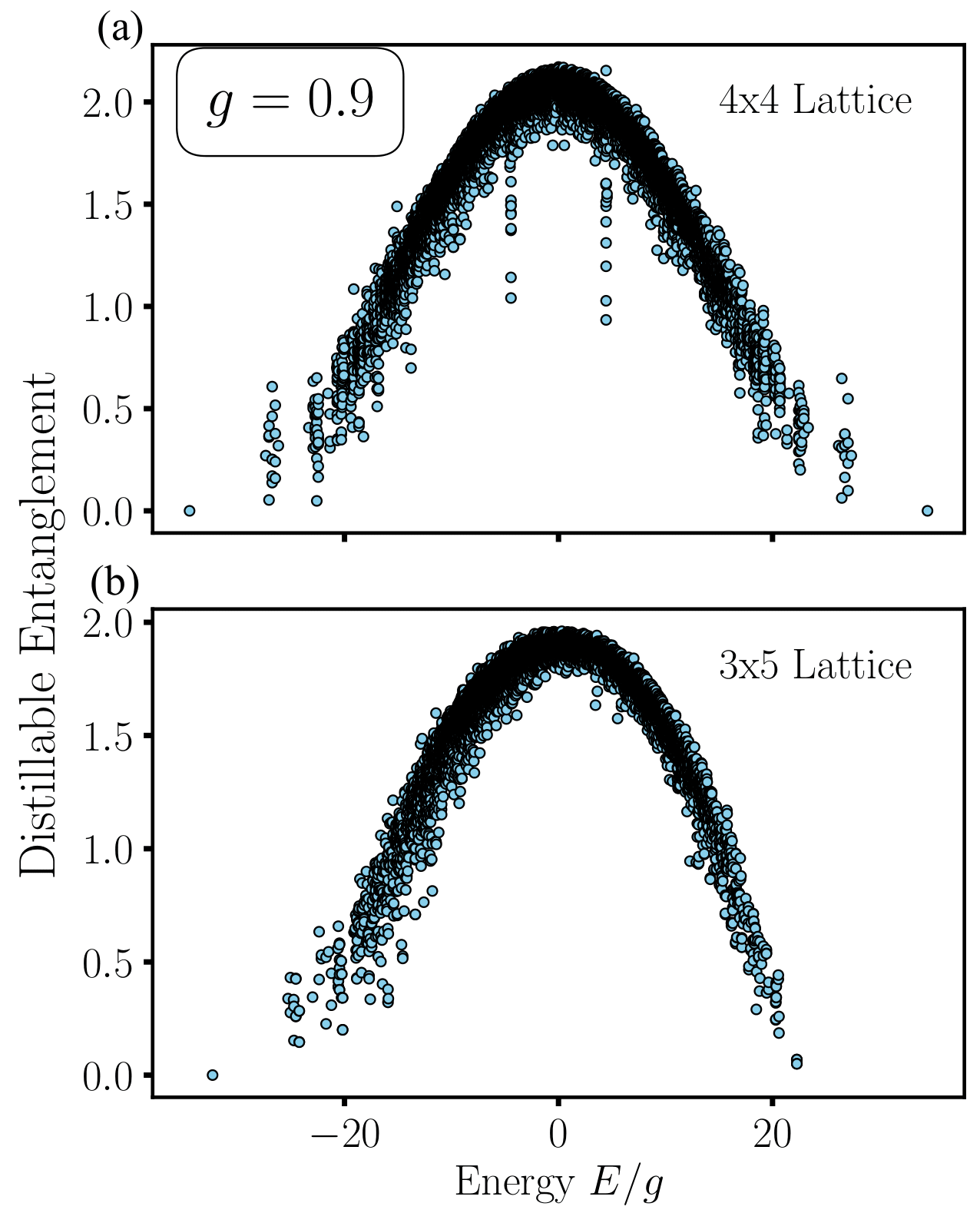}
    \caption{\textit{Distillable Entanglement of eigenstates for lattice geometries  $L\times k$, where $k>2$.} ((a) Upper Panel): No low entanglement states are observed for $4\times 4$ (here shown for $g=0.9$), and for ((b) Lower Panel):  $5\times 3$. While our analytical construction clearly does not extend to these geometries, this numerical study does not strictly rule out the possibility of different types of QMBS in these geometries, for instance potentially appearing only near specific coupling values.}
    \label{fig:large_lattice}
\end{figure}%
As discussed in the conclusion of the main letter, our analytical construction does not extend beyond the $L \times 2$ geometry. The specific scar states we identified are absent in larger geometries such as $L \times 3$ and $L \times 4$. Furthermore, we have explored these cases numerically for a limited range of coupling values $g$ and lattice sizes because we are constrained by the exponentially increasing computational cost. One example is shown in \Fig{fig:large_lattice}. We have found no evidence of low entanglement states, but cannot strictly rule out the potential existence of other, qualitatively different types of QMBS in these geometries.

\begin{figure}[t]
    \centering
    \includegraphics[width=0.92\linewidth]{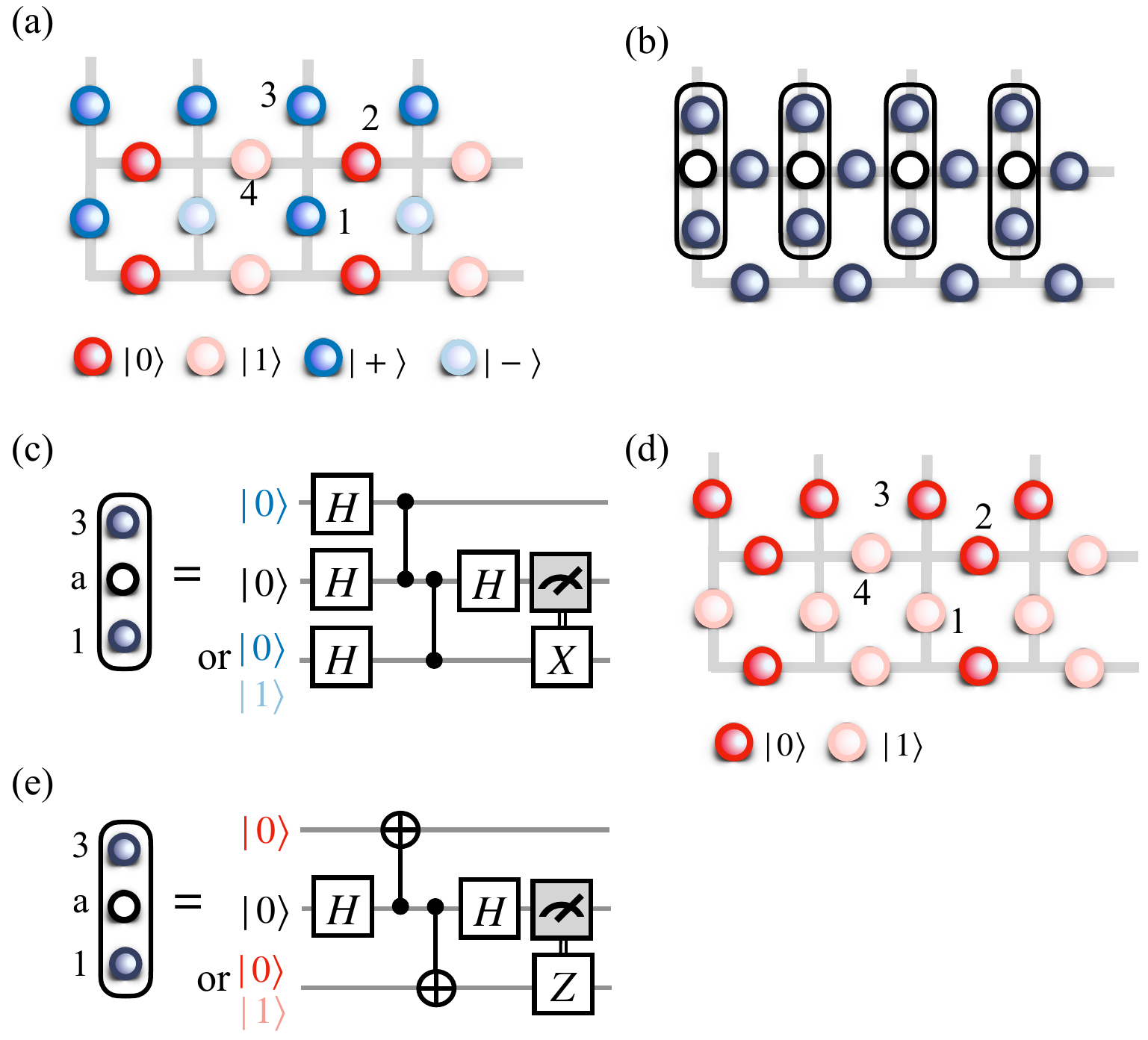}
    \caption{\textit{Alternative Circuit for QMBS Preparation using mid-circuit measurements.} (a) Initial state for the preparation algorithm in the LGT formulation ($V_y = 1$). Horizontal links are in $\sigma^z$ (electric) eigenstates, and vertical links in $\sigma^x$ (magnetic) eigenstates. Alternating $|0\rangle$ and $|1\rangle$ (or $|+\rangle$ and $|-\rangle$) yields eigenstates of any combination of stabilizers $Z_p Z_{p+L}$ and $X_p X_{p+L}$. (b) Mid-circuit measurement-based scheme for simultaneous  Gauss law eigenstate preparation using $L$ ancillas. (c) The circuit measures the combined $\sigma^z$ parity of two neighboring vertical links via controlled-$\sigma^z$ ($\Gamma(\sigma^z)$) operations and an ancilla {a}. The outcome dictates controlled-NOT operations on a vertical qubit, correcting Gauss laws in upper and lower rows, simultaneously. Control instructions classically depend on the desired target scar (via the horizontal links' state) and $V_y$ sector. 
    (d) A different starting point is shown for the scar state-preparation circuit,  a simultaneous eigenstate of the stabilizer $Z_p Z_{p+L}$ and the Gauss law constraints. (e) This initial states is not eigenstates of an $X_p X_{p+L}$; to transform it into asimultaneous eigenstate, a circuit is used involving mid-circuit measurements and $L$ ancillas. In a single shot, it measures $\sigma^x_{\nu}\sigma^x_{\nu'}$ values, where $\nu$ and $\nu'$ denote vertically stacked, vertically oriented links. From this the value of all $X_p X_{p+L}$ is determined.  Based on the measurement outcome and the desired target state, a controlled $\sigma^z$ operation is applied.
    \label{fig:stateprealtern}}
\end{figure}

\subsection{Details of the Experimental Realization}\label{app:detailsexpt}
In this section of the Supplemental Material, we detail the state-preparation circuits used to generate scar basis states, and discuss additional details. Time evolution is implemented using standard Trotterized circuits, which are well-documented in the literature for this model and are not discussed here. 

The experimental procedure discussed in the main text, for odd $L$, involves preparing a scar-subspace basis state versus a state outside this subspace, followed by time evolution. To prepare a scar basis state in the LGT a simple circuit, shown in \Fig{fig:stateprep} of the main text, is employed. The circuit prepares simultanoues  $X_pX_{p+L} = \pm 1$ (represented by $\prod_{\nu \in p}\sigma^x_\nu \prod_{\nu \in p+L}\sigma^x_\nu$ in the LGT),  $Z_pZ_{p+L} = \pm 1$ (corresponding to $\sigma^z_\nu$ in the LGT) and Gauss law eigenstates, with eigenvalues chosen according to any of the $4L$ basis states.  The circuit does not introduce horizontal entanglement between qubits, showing the rather trivial entanglement structure of the QMBS basis states; however, what is noteworthy is that any superposition of these states maintains zero distillable entanglement under time evolution.

An alternative approach, illustrated in  \Fig{fig:stateprealtern}(a-e), uses mid-circuit measurements~\cite{pino2021demonstration,rudinger2022characterizing,botelho2022error} with constant-depth~\cite{briegel2009measurement,aguado2008creation,tantivasadakarn2024long,baumer2024measurement}. This method is not strictly necessary: Although the model's ground state, the toric code, is long-range entangled for $g< g_c$, the scar subspace basis states that span the QMBS eigenstates are simple 2-qubit product states at any coupling (though the QMBS themselves are not). The circuit illustrated in panels (a-c) begins from a state that is an eigenstate of both $Z_p Z_{p+L}$ and  $X_p X_{p+L}$, but not of the Gauss law operators. Conversely, the circuit shown in panels (d-e) starts from states that are simultaneous eigenstates of both $Z_p Z_{p+L}$ and the Gauss laws, though not of $X_p X_{p+L}$. As in the original method, no entanglement is introduced along the horizontal direction of the lattice, resulting in the same final state.

\begin{figure}[t]
    \centering
    \includegraphics[width=0.85\linewidth]{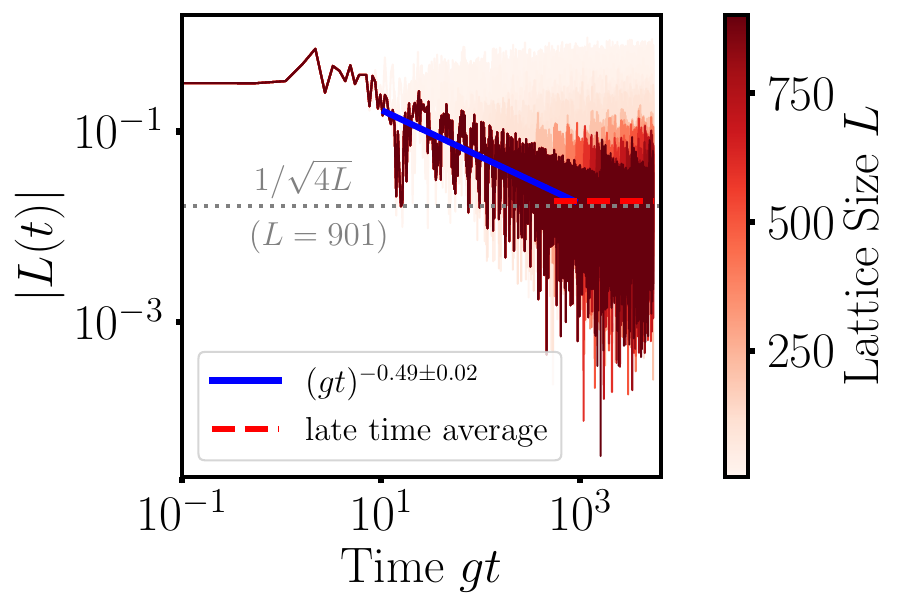}
    \caption{\textit{Loschmidt echo in the scar subspace for for large system sizes ${L}$.} The plot shows the absolute value of the Loschmidt echo, \( |L(t)| \), evaluated for scar initial states with system sizes up to $ L = 901 $. The echo exhibits a power-law decay, $\sim (gt)^{-1/2} $, approaching an asymptotic value of $ 1 / \sqrt{4L} $ at long times. The fit error is determined by varying the fit range and $L$.
    \label{fig:loschmidtscars}}
\end{figure}

To create a state outside the scar subspace, a $z$-product state that is also a Gauss law eigenstate is initialized. This state will naturally have well-defined $Z_p Z_{p+L} = \pm 1$ eigenvalues. To ensure it has no non-zero overlap with the scar subspace, the $Z_p Z_{p+L} = \pm 1$ pattern is carefully chosen.

In this manuscript, we focus on the Loschmidt echo—defined as the overlap between the initial and time-evolved states—and the entanglement of the time-evolved states as key observables. The Loschmidt echo, $L(t) $, is a widely used metric to characterize the return probability of time-evolved states and can be measured via an interferometric scheme. This approach involves preparing an ancilla in a Hadamard superposition, using it to control the time-evolution operator, and performing measurements in the $ x$- and $y$- bases to obtain the real and imaginary components of $ L(t) $, respectively~\cite{somma2002simulating,knap2013probing,mueller2023quantum}. In our analysis, we consider the behavior of $| L(t)|$ at asymptotic times. On a digital quantum computer this would require Trotterization. Using large Trotter steps risks artificially enhancing the return probability by causing unphysical recurrence to the initial state. Thus very deep circuits are required, which is challenging for near-term devices. Alternatively, instead of its asymptote, one may consider the decay of $|L(t)|$ at early times which is behaves as $1/\sqrt{gt}$ for an initial state within the scar subspace as shown in \Fig{fig:loschmidtscars} for $L=901$. Computing the same for sufficiently large $L$, is classically intractable for a non-scar initial state and is thus not shown here. However, the Loschmidt echo of a thermalizing system  should decay exponentially~\cite{googlescholarlohschmit}.

The distillable entanglement of time-evolved states can be measured directly using symmetry-conscious $k$-designs that have been proposed in~\cite{bringewatt2024randomized}, including for the model that is considered here. This scheme enables the measurement of all required quantities such as  $p_s = \text{Tr}(\rho_A^{(s)})$, as well as $k$-fidelities, $\text{Tr}( ( \rho_A^{(s)})^k)$.  From these, the distillable von Neumann entropy can be computed using $S^{(s)}_{\rm vN} = -\lim_{k\rightarrow 1+} \frac{\rm d}{{\rm d}k} \text{Tr}  ((\rho_A^{(s)})^k )$
for sufficiently large $k$, typically $k\approx 3-4$ is sufficient~\cite{bringewatt2024randomized}. 
\begin{figure}[t]
    \centering
\includegraphics[width=0.62\linewidth]{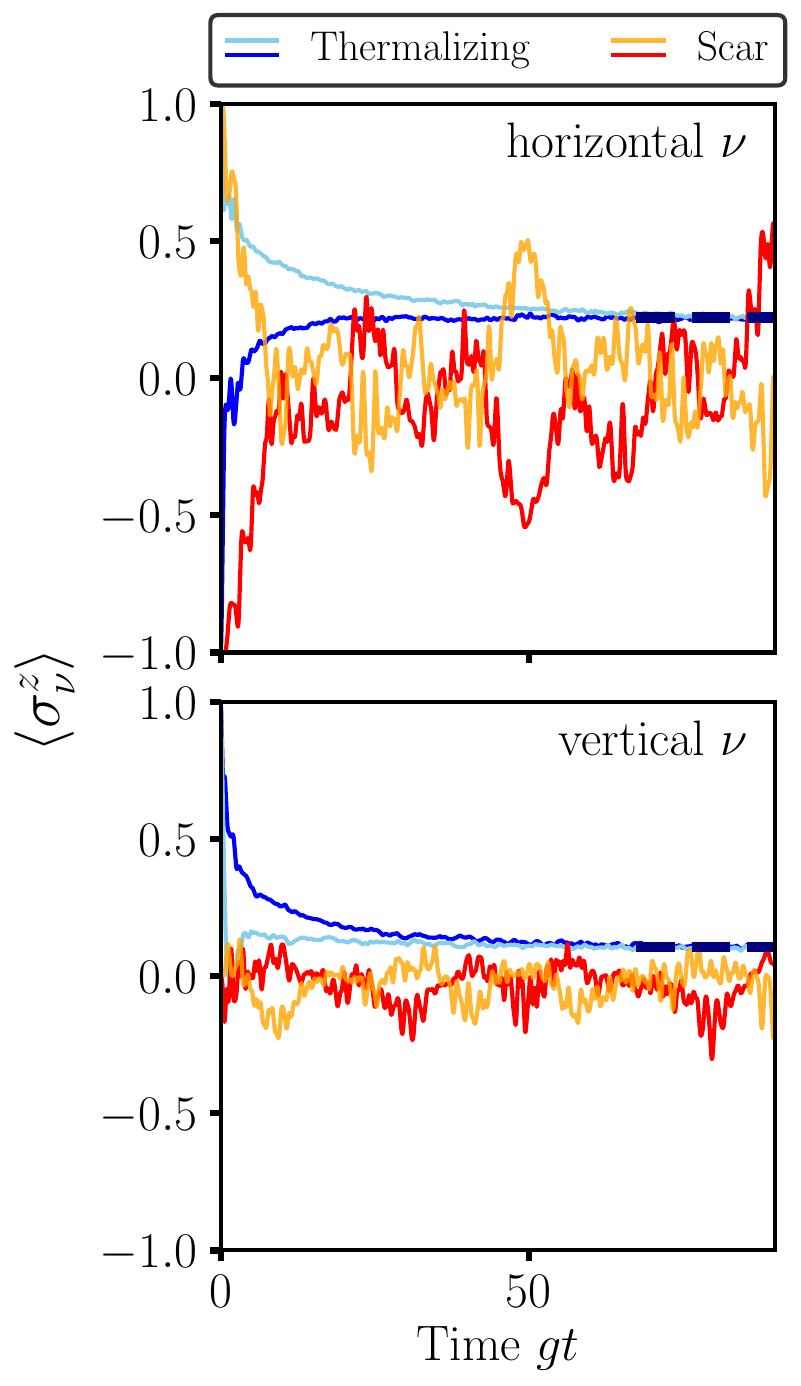}
    \caption{\textit{Local Observables.} The expectation values of electric operators, $\langle \sigma^z_\nu \rangle$, d for links oriented horizontally (top panel) and vertically (bottom panel), contrasting scar initial states (red and orange lines) with thermalizing non-scar initial states (dark and light blue lines) for $L=9$ and $g=0.9$. For non-scar initial states, the expectation values thermalize to a constant, asymptotic value (indicated by dashed blue lines). In contrast, the expectation values within the scar subspace exhibit oscillatory behavior, albeit without perfect revivals, a consequence of the non-equidistant energy spacing of scar eigenstates. This limits the utility of local observables, as distinguishing these oscillations from finite-volume effects and experimental imperfections may prove challenging. We argue that observables such as the Loschmidt echo or the distillable entanglement, however, allow a stronger distinction between scar and non-scar dynamics.
    \label{fig:localobs}}
\end{figure}
\begin{figure}[t]
    \centering
\includegraphics[width=0.68\linewidth]{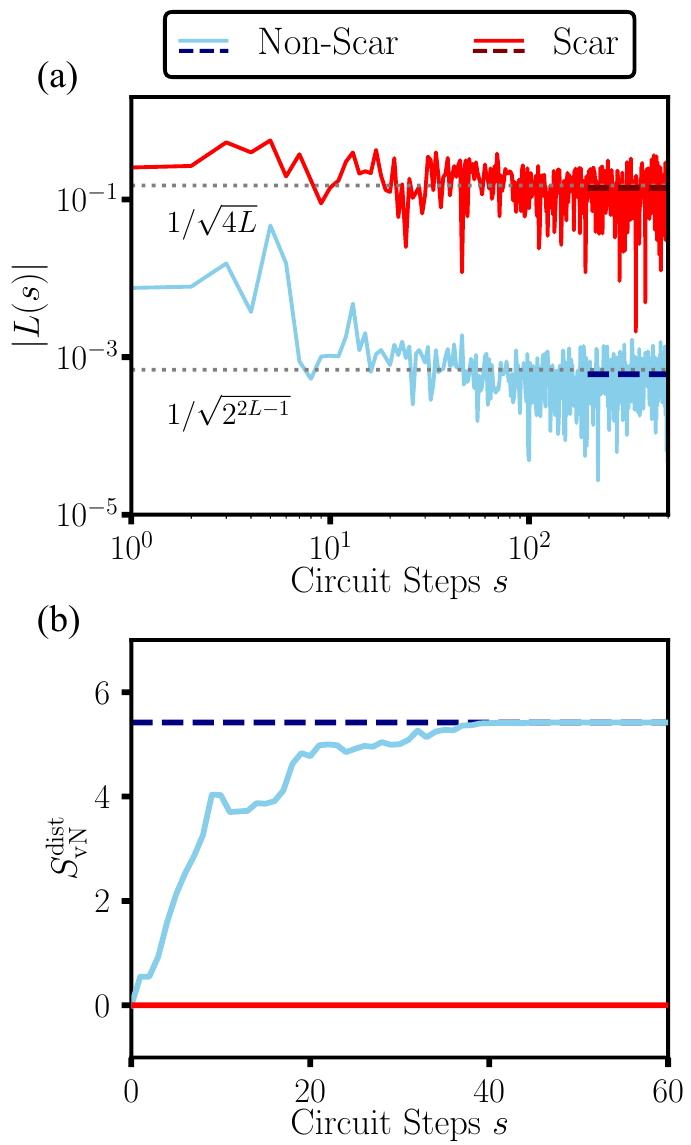}
    \caption{\textit{Experimental Realization with a Random Circuit.}
(a) Evolution of scar and non-scar states through random circuits, as defined in \Eq{eq:rand}. The absolute value of the Loschmidt echo, $ |L(s)| $, is shown as a function of the number of circuit steps $ s $, contrasting an initial state within the scar subspace (red lines) with a state outside this subspace (blue lines) for a system size $ L = 11 $. Dashed lines represents late-time averages. For the scar state, the echo decays to $ |L(s)| \sim 1/\sqrt{4L} $ (gray dotted line), where $ 4L$ is the dimension of the scar subspace. Conversely, the non-scar state decays to $|L(s)| \sim 1/\sqrt{2^{2L - 1}} $ (gray dotted line), reflecting the lack of an energy conservation constraint. 
(b) The distillable entanglement, $ S^{\rm dist}_{\rm vN} $, for scar versus non-scar initial states exhibits similar behavior to that shown in the main text for Hamiltonian evolution.
    \label{fig:echo_rand}}
\end{figure}

\subsection{Local Observables}
In previous QMBS studies, a more common observable are expectation values of local operators.
 \Fig{fig:localobs} shows the behavior of local observables for both scar-subspace initial states and thermalizing states. Specifically, we consider the expectation value of electric operators, $\langle \sigma^z_\nu\rangle$, for horizontal (top panel) and vertical links (bottom panel), comparing scar dynamics (red and orange lines) versus non-scar dynamics (blue and light blue curves). Each panel shows two curves corresponding to horizontal and vertical links at distinct lattice locations. Notably, while scar states exhibit somewhat more pronounced oscillations compared to thermalizing states—where local observables tend toward stationary values—no significant recurrences are observed in the local observables for scar-initial states. This lack of recurrence arises from the non-uniform energy spacing of the scar eigenstates. Because oscillations of similar magnitude may arise from finite-volume effects or experimental imperfections, for instance slight gate over/under rotations vary the value of $g$ over space and time and additionally smoothen the curves,  we believe that local observables offer less contrast between scar and non-scar dynamics than observables like the Loschmidt echo and distillable entanglement.
Future work will include a stability analysis of the scar solutions against experimental imperfections and Trotter evolution.

\subsection{Random Circuit Evolution}
\label{app:random_circuit}
The protocol discussed in the main text  compares Hamiltonian evolution within the scar subspace to a thermalizing state outside of it. While valuable from a physics perspective for studying ergodic versus non-ergodic dynamics,  it involves Trotterization, which requires very deep circuits to achieve thermalization. To demonstrate a quantum advantage, an alternative is based on shallower randomized circuits. Towards this end, we use the following circuit,
\begin{align}\label{eq:rand}
    \mathcal{U} = \prod_{i=1}^s U(\boldsymbol{\alpha}_{i},\boldsymbol{\beta}_i,\boldsymbol{\gamma}_i)\,,
\end{align}
where $s$ is the number of circuit layers, and
\begin{align}
    U&(\boldsymbol{\alpha}_i,\boldsymbol{\beta}_i,\boldsymbol{\gamma}_i) \equiv 
    \exp\{ - \sum_{p=1}^L {\alpha_{i}(p)} Z_pZ_{p+L} \}\nonumber\\
   & \times \exp\{- \sum_{p=1}^L\beta_{i}(p) [X_p+X_{p+L}] \} 
    \nonumber\\
    &\times \exp\{-\sum_{p=1}^L {\gamma_{i}(p)}  [Z_p Z_{p+1} + Z_{p+L} Z_{p+L+1} ] \}\,.
\end{align}
The angles $(\boldsymbol{\alpha}_{i},\boldsymbol{\beta}_i,\boldsymbol{\gamma}_i)\equiv (\{ \alpha_{i}(p)\}\,, \{ \beta_{i}(p)\},\, \{ \gamma_{i}(p)\})$ are randomly chosen for every layer, thus the evolution is not constrained by energy conservation as in the main text.
In the following we choose $ \alpha_{i}(p)=\alpha_{i} $, $ \beta_{i}(p) = \beta_{i}$, and $ \gamma_{i}(p)= \gamma_{i}$, i.e., independent of $p$, and draw $(\alpha_i,\beta_i,\gamma_i)$ in every layer from a single-qubit circular unitary ensemble. Other randomizations are also possible.

As in the main text, initial states within the scar subspace remain confined to that subspace under randomized evolution, while states outside it, no longer constrained by energy conservation, explore the full Hilbert space. In~\Fig{fig:echo_rand}, we present the same observables as in our main study: the Loschmidt echo in (a) and the distillable entanglement in (b), both as functions of circuit steps $s$. The stark contrast between scar and non-scar dynamics is clearly evident.

 We note, however, that many randomized circuits may be classically simulable (see, e.g., \cite{angrisani2024classically}) and  their computational hardness is an area of active investigation. We believe, therefore, that the thermalization example in the main text thus provides a stronger case for a demonstration of a quantum advantage.

\end{document}